\documentstyle[preprint,revtex,eqsecnum]{aps}
\begin{document}
\draft
\preprint{}
\begin{title}
Many-body exchange-correlation effects in the lowest subband \\
of semiconductor quantum wires
\end{title}
\author{Ben Yu-Kuang Hu and S.\ Das Sarma}
\begin{instit}
Department of Physics, University of Maryland,
College Park, Maryland 20742-4111.
\end{instit}

\begin{abstract}
We consider theoretically the electron--electron interaction induced
exchange-correlation effects in the lowest subband of a
quasi-one-dimensional GaAs quantum wire structures.  We calculate,
within the leading order dynamical screening approximation ({\em i.e.}
the so-called $GW$ approximation of the electron gas theory), the electron
self-energy, spectral function, momentum distribution function,
inelastic scattering rate, band gap renormalization, and, the many-body
renormalization factor both at zero and finite temperatures, and, both
with and without impurity scattering effects.  We also calculate the
effects of finite temperatures and finite impurity scattering on the
many-body properties.  We propose the possibility of a hot-electron
transistor device with a large negative differential resistance which is
based on the sudden onset of plasmon emission by energetic ballistic
electrons in one dimension.  The issue of the existence or non-existence
of the Fermi surface among the interacting one-dimensional quantum wire
electrons is critically discussed based on our numerical results.
\end{abstract}

\pacs{PACS numbers:  73.20.Dx, 73.20.Mf, 71.45.Gm, 72.20.Ht}

\narrowtext
\section{INTRODUCTION}
\label{sec: intro}

Recently, there has been a great deal of interest\cite{reed} in ultranarrow
confined semiconductor systems, called quantum wire structures, where
the motion of the electrons is essentially restricted to be one-dimensional.
These systems are usually fabricated on high-quality two-dimensional
electron systems by confining the electrons in one of the remaining free
directions through ultrafine nanolithographic
patterning.\cite{exp,cingolani,plaut,goni,pinczuk}
In addition, other methods for fabricating wires, such as growth of wires
on tilted superlattices and vicinal substrates\cite{tilted}
have also been reported.
These developments have opened up possibilities for novel and exciting
experiments, and stimulated considerable theoretical
activity.\cite{kane,hu,ogawa,lee,huappl,hutobep}
At present, quantum wires with active widths as small as approximately
${300\;{\rm \AA}}$ and of negligible thickness have been fabricated,
with further
reductions in size and improvement in quality in the offing.  Quantum
wires have generated much interest both for the potential for practical
applications in solid-state devices such as high-speed transistors,
efficient photodetectors and lasers,\cite{sakaki,vinterbook}
and because they have afforded us
for the first time an experimental opportunity to study {\em real}
one-dimensional Fermi gases in a relatively controlled manner (much in
the same way that, in the past two decades or so, semiconductor
inversion layers, heterojunctions and quantum wells have generated
considerable activity in pure and applied research of the two-dimensional
electron gas).  Thus, both from the fundamental and applied physics
viewpoints, there is interest in understanding the electronic
properties of quasi-one-dimensional quantum wires.

Much of the fundamental theoretical understanding of electrons in
one-dimensional systems have come from work on the Tomonaga-Luttinger
model.\cite{tomlut,larkin,mahan}
The Tomonaga-Luttinger model makes some drastic simplifying assumptions
which allow one to solve the interacting problem completely.
One surprising result that is obtained from the solution of this model
is that even the smallest interaction results in a disappearance of the
Fermi surface, leading to a system which is non-Fermi liquid (in the
sense that the elementary excitations are very different from those
of the non-interacting system).
Therefore, one would expect that the experimental properties of the
semiconductor quantum wires should be very different from any
predictions based on the assumption that the one-dimensional electron
gases are Fermi liquids.

Recently, quantum wires have been fabricated\cite{plaut,goni,pinczuk} in
which the one-dimensional quantum limit has been attained,
in the sense that only one
quantum subband is populated by the electrons in the quantum wire,
so that a one-dimensional interacting Fermi gas model is valid.
Contrary to what one might expect based on the well-established
theoretical results, the experimental data obtained from these experiments
on these structures, such as Raman scattering and photoluminescence,
can be successfully
explained on the basis of standard Fermi liquid theory.  This fact is also
mirrored by the success of other theories, such as the Landauer-B\"uttiker
formula,\cite{landauer} which treat electrons in mesoscopic
quasi-one-dimensional systems as noninteracting entities.
Thus we are posed with the question, ``Why are experimental results
from {\em real} quantum wires explicable on the basis of Fermi liquid
theories?"  This is one of the points we attempt to shed light on in
this paper.  We show that the interplay between plasmons
and impurities play an important role in determining the physical
characteristics of the system.  In particular, we show that the
physical mechanism responsible for the disappearance of a Fermi surface
in a clean system is the {\em virtual plasmon emissions}, and that in
an impure system these virtual plasmons emissions are damped, leading
to a {\em restoration} of the Fermi surface.  (The plasmons in our model
are equivalent to the Tomonaga bosons of the Tomonaga-Luttinger model.)

We also present calculations for experimentally measurable
properties for quantum wires.  For the quantum wire to be in the quantum
limit, the doping must necessarily be low
$\alt 10^{6}\,{\rm cm}^{-1}$
(otherwise, higher subbands are filled) and hence the Fermi energy and
other energy scales pertinent to the system are also small.  Therefore,
temperature effects are important, even at relatively low temperatures.
We have therefore presented the calculations for both $T=0$ and finite
temperatures.  To perform the $T\ne 0$ calculations, we have
developed a formalism for the efficient calculation of the
finite-temperature self-energy (valid for arbitrary dimensionality),
which we describe in this paper.

The many body properties of one-dimensional systems have unique
properties, one of which is the singular behavior of the mean free
path as a function of electron energy, which occurs when electrons
cross the plasmon emission threshold.   We discuss the possibility
that this singular behavior could in principle be applied to
produce a device with large negative differential resistivity.
Finally, since the the one-dimensional plasmons are
prominent protagonists in the physics of interacting one-dimensional
system, we conclude the paper with a discussion of the plasmons
in the presence of impurities and at finite temperature.

This paper is organized as follows.  In Sec.\ \ref{sec:sec2},
we study the many-body
properties of quantum wires at $T=0$.  In Sec.\ \ref{sec:sec3},
we develop the
formalism (for arbitrary dimensionality) necessary for efficient
calculation of the self-energy at finite temperatures, which we apply
to calculate the many-body properties of a quantum wire. In
Sec.\ \ref{sec:sec4},
we discuss the possible device applications of quantum wires.
In Sec.\ \ref{sec:sec5}, we discuss the plasmon spectrum of quantum wires, in
clean and dirty systems, and at finite temperature.
Sec.\ \ref{sec:sec6} contains
our summary and conclusions.  Some of the results in this papers have
been published previously in brief communications.\cite{hu,huappl,hutobep}
Also note that in this paper, $\hbar =1$ and $k_{\rm B} = 1$.

\section{MANY-BODY PROPERTIES OF A QUANTUM WIRE AT $T=0$}
\label{sec:sec2}

In this section, we present calculations of the many-body properties of a
one-dimensional quantum wire, using the so-called $GW$ approximation.
We begin with a brief review of the formalism, and then we describe
how we put effects of impurities into the system.  We then give
results and conclude this section with a discussion.

We assume that the quantum wire is formed by confining the
electrons to the two-dimensional plane $x$--$y$ plane (by,
say, modulation doping an AlGaAs-GaAs-AlGaAs heterostructure),
and then further confining the electrons
along the $y$-direction.   We assume that the confinement in the
$z$-direction is much stronger than the confinement in the
$y$-direction,  which is a reasonable approximation because,
at this stage, the technology for
confining electrons to two-dimensions is much more advanced
the technology for confining the electrons along an additional direction.
For example, by modulation doping, the electrons can be confined
in the $z$-direction on the order of less than $100\;{\rm \AA}$, whereas
currently the best confinement in the $y$-direction is approximately
$300\;{\rm \AA}$, leading to at least an order of magnitude difference
in the energy level spacings of the $y$- and $z$- directions.
Thus, in this paper, we assume that the electron gas has zero thickness
in the $z$-direction, but has a finite width in the $y$-direction.
(Note that it is necessary to assume a nonzero width so that the
one-dimensional Coulomb interaction is finite.\cite{lai})
The confinement of the electrons in the $y$--$z$ plane leads to
the quantization of energy levels into different subbands, depending
on what the wavefunction in the $y$--$z$ plane is.
We assume that the energy separation between the
lowest energy and higher energy subbands is so much larger than all
other energy scales in the problem that the higher subbands can be ignored.
Therefore, all quantities in the main part of this paper
refer to the lowest energy subband.
In Appendix \ref{app:appa}, we give the formalism
in the case of multiple subbands and briefly discuss why,
for processes with energies much less than the subband energy
separation, the higher subbands have negligible contribution.

\subsection{The Hamiltonian}
\label{subsec:hamilt}

The Hamiltonian of the system is given by
$\hat H = \hat H_0 + \hat H_{\rm e-e} $
where $\hat H_0$ is the bare kinetic energy term
and $\hat H_{\rm e-e}$ is the electron--electron interaction term,
given by
\begin{eqnarray}
\hat H_0 &=& \sum_{k\sigma} \xi_(k)\;
\hat c^\dagger_{k \sigma} \hat c_{k \sigma}\nonumber
\\& & \\
\hat H_{\rm e-e} &=& {1\over 2 L}
\sum_{kk'q}\sum_{\sigma\sigma'}
V_c(q)\;
\hat c^\dagger_{k+q,\sigma}\,\hat c^\dagger_{k'-q,\sigma'}
\,\hat c_{k' \sigma'}\,\hat c_{k \sigma},\nonumber
\end{eqnarray}
Here $\hat c_{k \sigma}^\dagger$ ($\hat c_{k \sigma}$)
are fermion creation (annihilation) operators for
for states with wavevector $k$ in the
$x$-direction and electron spin $\sigma$,
$\xi(k)$ is the is the kinetic energy relative to the
chemical potential $\mu$, and $L$ is the length of the system.
The $V_c(q)$ are the matrix elements of the Coulomb interaction
(screened by the static lattice dielectric constant $\epsilon_0$),
\begin{eqnarray}
V_c(q) &= &
\int_{-\infty}^{\infty}\! dy \int_{-\infty}^\infty\! dy'
\int_{-\infty}^{\infty}\! dx\ e^{-iqx}
{e^2\over\epsilon_0\Bigl[(x-x')^2 + (y-y')^2\Bigr]^{1/2}}
|\phi(y)|^2 \, |\phi(y')|^2\nonumber
\\ & & \\
&=& \int_{-\infty}^{\infty} dy\int_{-\infty}^{\infty} dy'\ v(q,y-y')
|\phi(y)|^2 \, |\phi(y')|^2,\nonumber
\end{eqnarray}
where $\phi(y)$ is the transverse wavefunction and
\begin{equation}
v(q,y-y') = {2 e^2\over \epsilon_0} K_0(|q(y-y')|)\label{vqyy}
\end{equation}
is the one-dimensional Fourier transformation of the Coulomb
interaction.\cite{lai,li}  Here, $K_0(x)$ is the zeroth-order
modified Bessel function of the second kind.\cite{abram}

For an external confining potential that is parabolic,
the self-consistent confining potential (that
is, the external confining potential plus the self-consistent
Hartree term) gives approximately a square well potential.\cite{kumar}
This turns out to be a reasonable approximation for modeling quantum
wire confinement.
Therefore, we approximate the self-consistent potential by a
square well with infinite barriers at $y=-a/2$ to $y=a/2$, so that
\begin{equation}
\phi(y) = \left\{
\begin{array}{ll}
\displaystyle{2\over a} \cos\biggl({\pi y\over a}\biggr),
&\mbox{if $-a/2\le y\le a/2$};\\ \\
0,& \mbox{otherwise};
\end{array}\right.
\end{equation}
and the Coulomb matrix element in this square-well case is
\begin{eqnarray}
V_c(q)
&=& {2 e^2\over \epsilon_0}\Bigl[{2\over a}\Bigr]^2
\int_{-a/2}^{a/2} dy\int_{-a/2}^{a/2} dy'\;
K_0(|q(y-y')|) \cos^2\Bigl({\pi y\over a}\Bigr)
               \cos^2\Bigl({\pi y'\over a}\Bigr)\nonumber
\\ & & \label{defvcq}\\
&=& {2 e^2\over \epsilon_0}
 \int_{0}^{1} dx\; K_0(|qa|x)\,\Bigl[(2-(1-x)\cos(2\pi x) +
{3\over 2\pi} \sin(2\pi x)\Bigr]\nonumber
\end{eqnarray}
whose asymptotic forms are
\begin{equation}
V_c(q) =
\left\{ \begin{array}{ll}
\displaystyle{3\pi e^2\over\epsilon_0 |qa|}, &
\mbox{for $|qa|\rightarrow\infty$;}\\ & \\
\displaystyle{2 e^2\over\epsilon_0} [K_0(|qa|) + 1.9726917...],\ \
			  &\mbox{for $|qa|\rightarrow 0$.}
\end{array}\right.
\end{equation}
Note that $K_0(x) \sim -\ln(x)$ as $x\rightarrow 0$, so that
the Coulomb matrix element diverges as $-2e^2 \ln(|qa|)/\epsilon_0$
small $q$.  To speed up numerical computation,
we fitted $V_c(q)$ with a function
which extrapolated between the $q\rightarrow 0$ and $q\rightarrow\infty$
forms and which deviated from the true function by no more than 0.6\%.

Throughout this paper, we assume that the band for the electrons in
the unconfined $x$-direction is parabolic, so that
$\xi(k) = \hbar^2 k^2 /(2 m^*) - \mu$, where $m^*$ is the
effective electron band mass.  This is an excellent assumption for
GaAs quantum wires because
the relevant electron densities are so low ($k_F\sim 10^6\,{\rm cm}^{-1}$)
that the nonparabolicity in the band for $k
\alt k_F$ is insignificant.

\subsection{The electron self-energy $\Sigma(k,\omega)$ within the GW and RPA
approximation}
\label{subsec:elecse}

A quantity which provides a substantial amount of information on an interacting
many-electron system is the electron Green's function $G(k,\omega)$,
or equivalently, the electron self-energy, $\Sigma(k,\omega) =
G_0^{-1}(k,\omega)-G^{-1}(k,\omega)$ (where $G_0(k,\omega)$ is the bare
noninteracting Green's function).
The self-energy is roughly the correction to the noninteracting electron
single-particle energy due to the interactions.
Once $G(k,\omega)$ or $\Sigma(k,\omega)$ is known, the
one-electron properties of a system, such as the
spectral density function, electron distribution function and
band-gap renormalizations, can be calculated.

Generally, it is impossible to calculate the
$\Sigma(k,\omega)$ exactly for interacting systems, and one must resort
to various approximation schemes.
However, in one dimension, the self-energy of the Tomonaga-Luttinger model,
which uses somewhat artificial assumptions
(specifically (1) two completely linearly dispersing bands of electrons
with an infinite bandwidth,
populated by an infinite density of electrons and (2) a short-ranged
interaction),
can be calculated exactly.\cite{tomlut,larkin}  As alluded to in the
introduction, this models predicts that any interaction, no matter
how small, will drive the system away from Fermi liquid behavior.
However, the connection between the exactly
solvable but artificial Tomonaga-Luttinger model and
real semiconductor quantum wires is somewhat tenuous, and
there is a need for calculations on
more realistic models which can be compared to experiment.
The model we employ assumes that the electron dispersion in the
free direction of the quantum wire is parabolic, and the electrons
interact via the exact Coulomb interaction.
Unfortunately, with these more realistic assumptions,
the problem is no longer exactly soluble and therefore,
as in higher dimensions, $\Sigma(k,\omega)$
can only be calculated approximately.
In this paper, we use the so-called $GW$ approximation to calculate the
self-energy,\cite{hedin} where the Green's function is dressed by
the dynamically screened Coulomb interaction, which is calculated within
the random phase approximation (RPA).
The Feynman diagram for the self-energy and the screened interaction
in these approximations are shown in Fig.\ \ref{fig1}(a) and \ref{fig1}(b).
This approximation, which sums up the largest diagrams in the
self-energy perturbation series, has long been employed to successfully
calculate properties of three and two dimensional electron
systems.\cite{twodim,ando}  We ignore diagrams which are higher order in the
screened interaction, such the diagram shown in Fig.\ \ref{fig1}(c), because we
expect them to have a smaller contribution.  In fact, in Appendix
\ref{app:appb}
we show that at $k=k_F$ and $\omega=\xi_{k_F}$, and for a short-ranged
interaction, Fig.\ \ref{fig1}(c) gives zero contribution.

The self-energy within the $GW$ approximation at $T=0$ is
\begin{equation}
\Sigma(k,\omega) = i\int_{-\infty}^\infty {dq\over 2\pi}
\int_{-\infty}^\infty {d\omega'\over 2\pi}\;
W(q,\omega') G_0(k-q,\omega-\omega').
\end{equation}
where, $G_0(k,\omega) = [w-\xi_k + i0^+{\rm sgn}(k-k_F)]^{-1}$ and
$W(q,\omega)$ is the screened Coulomb interaction, which given by
\begin{equation}
W(q,\omega) = {V_c(q)\over\epsilon(q,\omega)}.
\end{equation}
Here, $\epsilon(q,\omega)$ is the dielectric function, which describes
the screening properties of the electron gas.
The self-energy can be separated into the
frequency-independent exchange and correlation parts
\begin{equation}
\Sigma(k,\omega) \equiv \Sigma_{\rm ex}(k) + \Sigma_{\rm
cor}(k,\omega).
\end{equation}
The exchange part is given by\cite{mahan}
\begin{equation}
\Sigma_{\rm ex}(k) = -\int {dq\over 2\pi}\; n_F(k+q) V_c(q),
\end{equation}
where $n_F(k+q) = \theta(k_F -|k+q|)$ is the Fermi function at $T=0$,
and $\Sigma_{\rm cor}(k,\omega)$ is defined to be the part of
$\Sigma(k,\omega)$ not included in $\Sigma_{\rm ex}(k)$.
In the $GW$ approximation, the ${\Sigma_{\rm cor}(k,\omega)}$
can be written in the line and pole decomposition\cite{mahan,ferrell}
\begin{mathletters} \begin{eqnarray}
\Sigma_{\rm cor}(k,\omega) &=& \Sigma_{\rm line}(k,\omega)
+\Sigma_{\rm pole}(k,\omega),\\
\Sigma_{\rm line}(k,\omega) &=& -\int_{-\infty}^\infty {dq \over
2\pi} V_c(q) \int_{-\infty}^\infty {d\omega' \over 2\pi}
{1\over (\xi_{k+q} - \omega) - i\omega'}
\Bigl[{1 \over \epsilon(q,i\omega')}-1\Bigr],\\
\Sigma_{\rm pole}(k,\omega)
&=& \int_{-\infty}^\infty {dq\over 2\pi}
\Bigl[\theta(\omega-\xi_{k+q}) - \theta(-\xi_{k+q})\Bigr] V_c(q)
\Bigr({1\over\epsilon(q,\xi_{k+q}-\omega)}-1\Bigl).\label{imsigma}
\end{eqnarray}\end{mathletters}
The $\Sigma_{\rm  line}(k,\omega)$ is completely real
because $\epsilon(q,i\omega')$ is real and even with respect
to $\omega'$.  Thus, ${\rm Im}[\Sigma_{\rm pole}]$ gives the total contribution
to the imaginary part of the self-energy.
In evaluating the self-energies, we assume that the frequency
$\omega$ has an infinitesimal positive imaginary part;
this gives us the {\em retarded} self-energy.\cite{mahan}
(Unless otherwise stated, all real frequency self-energies in
this paper are retarded.)

Within the RPA, $\epsilon(q,\omega)$ is given by
$\epsilon(q,\omega) = 1 - V_c(q)\Pi_0(q,\omega),$
where $\Pi_0(q,\omega)$ is the irreducible polarizability, given
diagrammatically by the particle-hole bubble.
In a pure system,\cite{li,lai,williams}
\begin{equation}
\Pi_0(q,z) =  {m\over \pi q} \ln \Bigl[{z^2 - ((q^2/ 2m) - q v_F)^2
\over z^2 - ((q^2/ 2m) + q v_F)^2}\Bigr],\label{pizero}
\end{equation}
where the principal value of logarithm ($|{\rm Im}[\ln]| < \pi$)
should be taken.
In evaluating $\Pi_0(q,\omega)$ for real frequencies,  the limit
$z = \omega + i0^+$ should be taken.
The RPA form of $\epsilon(q,\omega)$ has recently been
shown\cite{qpli} to {\em exactly} reproduce the
plasmon dispersion of one-dimensional systems, in the sense that the
long wavelength RPA correctly gives the exact Tomonaga boson dispersion
of the Tomonaga-Luttinger model.

The $\Pi_0(q,\omega)$ given in Eq.\ (\ref{pizero}) assumes that the
system is free from any defects.
While modulation doping techniques result in quantum wires are
generally of high quality, impurities and other imperfections always
exist.  In general, the effect of these impurities is to
cause the electrons to diffuse instead of moving ballistically.
This diffusive motion of the electrons affects the polarizability
(since it is the response of the electrons to an applied potential)
and hence electronic screening, generally tending to
make screening less effective since electrons cannot move as
quickly to their screening positions.
Impurity effects are usually introduced diagrammatically into
the screening in the RPA by including impurity ladder diagrams into
the electron-hole bubble,\cite{agd} which yields
a polarizability $\Pi_\gamma(q,\omega)$ which
has a diffusive regime for small $q$ and $\omega$.
Since the exact expression
for $\Pi_\gamma(q,\omega)$ within this diagramatic approach is complicated,
we use a particle-conserving expression, given by Mermin,\cite{mermin} which
captures the essential physics of impurity collisions on
$\Pi_\gamma(q,\omega)$.
For an impurity scattering rate $\gamma$,
\begin{equation}
\Pi_\gamma(q,\omega) = {(\omega + i \gamma) \Pi_0(q,\omega +
i\gamma) \over \omega + i\gamma [\Pi_0(q,\omega + i \gamma)
/\Pi_0(q,0)]}.\label{mermin}
\end{equation}
The form given in Eq.\ (\ref{mermin}) has the correct diffusive behavior at
small $q$ and $\omega$, which is absent in the $\Pi_0(q,\omega)$ for
the pure case.
In particular, for $q,\omega\rightarrow 0$,
\begin{eqnarray}
\Pi_\gamma(q,\omega) &\approx& -{ 2 m_e\over \pi k_F \hbar^2}
{D q^2 \over D q^2 + i\omega},\nonumber\\ \\
\Pi_0(q,\omega) &\approx& {n q^2 \over m^* \omega^2}\qquad\qquad
({\rm for}\ q\ll \omega/v_F),\nonumber
\end{eqnarray}
where $D = v_F^2/\gamma$ is the one-dimensional diffusion constant.
The modification of the behavior of the polarizability due to impurities
has important consequences for long-wavelength plasmons, which
in turn affect the many-body properties of the one-dimensional
electron systems significantly.

For frequencies on the imaginary axis (as is the case in the integrand
of the line component of the self-energy), the polarizability is given by
\begin{equation}
\Pi_\gamma(q,i\omega') = {(|\omega'| +\gamma) \Pi_0(q,i|\omega'| +
i\gamma) \over |\omega'| + \gamma [\Pi_0(q,i|\omega'| + i \gamma)
/\Pi_0(q,0)]}.
\end{equation}

\subsection{Calculation and Results}
\label{sec:calc}

Using the expressions given above, we calculate the self-energy
for a quantum wire for both pure and impure cases.
We briefly discuss some subtleties involved in the numerical calculation,
and then we show our results.

Both the $\Sigma_{\rm ex}(k)$ and $\Sigma_{\rm line}(k,\omega)$
are comparatively straightforward to evaluate numerically.
However, evaluation of $\Sigma_{\rm pole}(k,\omega)$ for the clean case
($\gamma = 0$) is complicated by the singularities present in the integrand.

For ${\rm Im}[\Sigma_{\rm pole}(k,\omega)]$, in addition to the
contribution from single-particle excitations (${\rm Im}[\epsilon]\ne 0$),
there are also contributions from the plasmon excitations
(${\rm Re}[\epsilon] = 0$ and $|{\rm Im}[\epsilon]| = 0^+$),
which produce $\delta$-functions in the integrand of
${\rm Im}[\Sigma_{\rm pole}(k,\omega)]$ (see Fig.\ \ref{fig2}).  Care
must be taken to pick up this $\delta$-function in a numerical integration.
To numerically evaluate the plasmon contribution to ${\rm Im}[\Sigma]$,
we performed a search along $q$ for ${\rm Re}[\epsilon(q,\xi_{k+q}-\omega)]
= 0$, in the regions ${\rm Im}[\Pi_0(q,\xi_{k+q}-\omega)]=0$.
When a zero of ${\rm Re}[\epsilon]$ was found,
we evaluated the integral by two methods:
(1) introducing a small damping term $\gamma$ in
$\epsilon$ and numerically integrating around the ${\rm Re}[\epsilon]=0$ region
and (2) numerically evaluating
$\partial\epsilon(q,\xi_{k+q}-\omega)/\partial q$, which is proportional
to the weight of the $\delta$-function.  Both methods gave the same results.
Since the danger of missing the plasmon peak in the case of small $\gamma$ is
always present, we routinely searched for the ${\rm Re}[\epsilon]=0$
even when $\gamma\ne 0$.

In the integrand for ${\rm Re}[\Sigma_{\rm pole}(k,\omega)]$,
at the $q_0$ where ${\rm Re}[\epsilon] = 0$,
there is a $(q - q_0)^{-1}$ principal part
divergence, which also can cause numerical problems if it is not handled
carefully.  We treat this the problem either by cutting off the integrand
around $q_0$ when its absolute value exceeded a certain
value or by putting in a small damping term $\gamma$, and
paying special attention to the integration around $q_0$.
Again, both methods gave the same results.

{}From $\Sigma(k,\omega)$, we can obtain the single-particle spectral
function\cite{mahan,fetter}
\begin{equation}
A(k,\omega) = {2|{\rm Im}[\Sigma(k,\omega)|\over
(\omega-\xi_k - {\rm Re}[\Sigma(k,\omega)])^2 +
({\rm Im}[\Sigma(k,\omega)])^2.}\label{spectral}
\end{equation}
(Note that ${\rm Im}[\Sigma]$ is assumed to contain an infinitesimal
negative part.)
The spectral function $A(k,\omega)$ can roughly be interpreted as
the probability density of the different energy eigenstates required
to make up a particular $k$-state.  It satisfies the sum rule
\begin{equation}
\int_{-\infty}^\infty {d\omega\over 2\pi} A(k,\omega) = 1,
\label{asumrule}
\end{equation}
which is generally satisfied to within less than a percent in all
our numerical calculations.
We pay particular attention to the self-energy and spectral function
at $k=k_F$, since the behavior of of these functions at the Fermi
surface determine the low energy properties of the system.
The exact solution of the Luttinger model indicates that
interactions produce a non-Fermi liquid, the so-called Luttinger liquid.
We reproduce this
result in clean systems, and show that virtual plasmon emission is
responsible for this behavior.  We then show that impurities suppress
virtual plasmon emission, and the Fermi surface reappears.

A system is a Fermi liquid if it possesses a Fermi surface ({\em i.e.},
a discontinuity in $n_k$) whose presence is indicated by
a $\delta$-function in $A(k_F,\omega)$ at $\omega = 0$.
The existence of a $\delta(\omega)$ in $A(k_F,\omega)$
depends crucially on the behavior of ${\rm Im}[\Sigma(k_F,\omega)]$
as $\omega \rightarrow 0$.
If $|{\rm Im}[\Sigma(k_F,\omega)]|$ goes to zero
faster than $|\omega|$, and
\begin{equation}
\omega-\xi_{k_F}-{\rm Re} [\Sigma(k_F,\omega)]\approx Z_F^{-1}\omega,
\label{linearize}
\end{equation}
where $Z_F$ is a constant called the renormalization
factor,\cite{mahan} then as $\omega\rightarrow 0$,
$|{\rm Re}[\Sigma(k_F,\omega)|\gg|{\rm Im}[\Sigma]|$, and hence
\begin{eqnarray}
A(k_F,\omega)
&\approx& \lim_{\varepsilon\to0} {2\varepsilon \over
Z_F^{-1}\omega^2 + \varepsilon^2}\nonumber\\
&=&{2\pi\,Z_F}\,\delta(\omega).\label{deltafn}
\end{eqnarray}
Thus, if $|{\rm Im}[\Sigma(k_F,\omega)]|$ goes to zero
faster than $|\omega|$, then there is a discontinuity of magnitude
$Z_F$ in $n_k$ at $k_F$.
Since $Z_F$ is the linear coefficient of the expansion in $\omega$ of the
left hand side of Eq.\ (\ref{linearize}), it is given by
\begin{equation}
Z_F = \Bigl|1 - \Bigl[{\partial{\rm Re}[\Sigma(k_F,\omega)]
\over\partial\omega} \Bigr]_{\omega = 0}\Bigr|^{-1}.
\label{19}
\end{equation}

In contrast, if $|{\rm Im}[\Sigma(k_F,\omega)]|$
goes to zero slower than $|\omega|$, then the spectral function
$A(k_F,\omega)$ does not take the form Eq.\ (\ref{deltafn})
as $\omega\rightarrow
0$, and hence there is {\em no} $\delta$-function in
$A(k_F,\omega)$, implying that the system is {\em not} a Fermi
liquid.\cite{tomlut,larkin,marginal}
In general,
interacting three-dimensional systems with and without disorder,
and, interacting pure two-dimensional systems are Fermi
liquids.\cite{luttinger,highdim,giuliani0}
Through a study of ${\rm Im}[\Sigma(k_F,\omega)]$,
we show that in one dimension within the GW approximation,
the system is not a Fermi liquid in
the absence of impurity scattering, but becomes a Fermi liquid
in the {\em presence} of scattering.

The imaginary part of the self-energy in the $GW$ approximation,
from Eq.\ (\ref{imsigma}), can be written as
\begin{equation}
{\rm Im}[\Sigma(k,\omega)]= {1\over 2}\int_{-\infty}^\infty {dq\over 2\pi}
\Bigl[\theta(\omega-\xi_{k+q}) - \theta(-\xi_{k+q})\Bigr]
P(q,\xi_{k+q}-\omega),
\end{equation}
where $P(q,E) = 2 V_c(q){\rm Im}[\epsilon^{-1}(q,E)]$
which is the Born approximation transition
rate for a momentum change $q$ and energy change $E$.\cite{pines}
Thus, ${\rm Im}[\Sigma(k,\omega)]$ can be interpreted as the total
scattering rate for an electron with momentum $k$ and energy
$\omega$.  By setting $\omega = \xi_k$, the actual electron
kinetic energy, and one obtains the total Born
approximation scattering rate.  For $\omega\ne\xi_k$,
${\rm Im}[\Sigma(k,\omega)]$ corresponds to the total scattering rate
due to {\em virtual} transitions, since the transitions violate
energy conservation.

At low energies in two and three dimensions, single-particle scattering
is far more important than plasmon scattering
because the single-particle excitation spectrum is
gapless and the phase-space available for single-particle scattering
extends around the entire Fermi surface, whereas the plasmon
dispersion either rises quickly or has a gap at $q=0$.
Therefore, for small $\omega$, the major contribution to ${\rm
Im}[\Sigma(k_F,\omega)]$ in two and three dimensions
comes from {\em virtual single particle excitations}.
In contrast, in one dimension, the single-particle excitation spectrum
has a gap except at $|q| = 0,2k_F$, and the phase-space available for
single-particle scattering is severely restricted,
while the plasmon dispersion is gapless at $q=0$.
Hence, in one dimension, ${\rm Im}[\Sigma(k_F,\omega)]$ at small
$\omega$ is dominated not by virtual single-particle excitations but
by {\em the virtual excitation of plasmons}.
It is the domination of the plasmon excitations at low energies that
give interacting one-dimensional systems their unique non-Fermi liquid
behavior.

In the case of a clean quantum wire ($\gamma = 0$), as
$\omega\rightarrow 0$, the contribution to ${\rm Im}[\Sigma(k_F,\omega)]$
of the low energy plasmon excitations goes as $|\omega| \,
|\ln(|\omega|)|^{1/2}$,
whereas the contribution of the
single-particle excitations goes as
$\omega/(\ln|\omega|)^2$.
Thus the plasmon contribution dominates, and
since ${\rm Im}[\Sigma(k_F,\omega)]$ goes to zero slower than $|\omega|$,
the Fermi surface
does not exist (in agreement with Luttinger liquid theory).
The Fermi surface is smeared out to the extent that a
momentum-space discontinuity in the distribution function
does not exist because of the ease with which
the particles at the Fermi surface can emit virtual plasmons.
In contrast, in two dimensions, the contribution
to ${\rm Im}[\Sigma(k_F,\omega)]$ of low energy plasmons goes as $\omega^2$,
whereas that of the single-particle excitations goes as $\omega^2
|\ln(|\omega|)|$
and therefore in the two-dimensional case, the single-particle
excitations dominate, and since the self-energy goes to zero
faster than $|\omega|$, the two-dimensional interacting electron gas
is a Fermi liquid.

The inclusion of impurity scattering causes the electrons to
diffuse at long wavelengths which damps out the plasmons at small
$q$ (see section \ref{sec:sec5}).
Hence, the plasmon contribution to ${\rm Im}[\Sigma(k_F,\omega)]$
at small $|\omega|$ is removed, and the entire contribution comes
from single-particle excitations.
In one dimension, it modifies the behavior to
$|{\rm Im}[\Sigma(k_F,\omega)]| \sim \omega^2
|\ln(|\omega|)|^3$
as $|\omega| \rightarrow 0$, which implies that the Fermi surface
is restored.
This result indicates that the Fermi surface is resurrected in dirty systems
because the low energy virtual plasmon emission responsible for
its destruction in clean systems has been suppressed by impurity scattering.
In two dimensions, however, the situation is reversed.  The plasmon
contribution was not dominant in the first place, so its removal is
not significant.  However, inclusion of impurity scattering
enhances\cite{gof4prime}
the single particle scattering rate (because of phase space effects)
and it results in a $|{\rm Im}[\Sigma(k_F,\omega)]|\sim |\omega|$.
This agrees
with the notion that an interacting disordered two-dimensional system is
not a Fermi liquid.\cite{gof4prime}
Table \ref{table1} summarizes the behavior of $[\Sigma(k_F,\omega)]$ resulting
from various contributions in both the clean and dirty cases, in both
one and two dimensions.  The details of the calculations are given
in Appendices \ref{app:appc} and \ref{app:appd}.

In Fig.\ \ref{fig3}, we show the real and the imaginary
parts of the self-energy
for $k=0$ (bandedge) and $k=k_F$ (Fermi surface), for $E = 0,E_F$,
for $k_F a = 0.9$ and
$r_s = 4 e^2 /(\pi \epsilon_0 v_F) = 1.4$.\cite{define}
In the $k=k_F$, clean ($\gamma = 0$) case, the $A(k_F,\omega)$
has no $\delta$-function at $\omega=0$.  The spectral function goes
as $A(k_F,\omega)\sim
(|\omega|\,|\ln(|\omega|)|^{3/2})^{-1}.$
In the dirty case
$\gamma\ne 0$, we have $A(k_F,\omega)\sim|\ln(|\omega|)|^3
+ 2\pi Z_F\delta(\omega)$.  We can see by comparison of the $\gamma = 0$
and $\gamma = E_F$ lines in Figs.\ 2(c) that some of the spectral weight
around the origin in the $\gamma=0$ case has been transferred to
the $\delta$-function in the $\gamma\ne 0$ case.

In Fig.\ \ref{fig4}, we show our calculated Fermi
distribution function\cite{mahan}
\begin{equation}
n_k = \int_{-\infty}^0 {d\omega\over 2\pi}\,
A(k,\omega)\label{nkfroma}
\end{equation}
for various values of the impurity scattering
rate $\gamma$.  We emphasize that $\gamma$ was included only in the
dynamical screening function and {\em not} in the single-electron Green's
function because we wanted to determine if the suppression of
the emission of low energy virtual plasmons produces a discontinuity in
$n_k$. In the next subsection, we discuss
the effects of introducing disorder into the bare Green's function.
Fig.\ \ref{fig4}(a) clearly shows a discontinuity in $n_k$
at $k=k_F$ for $\gamma/E_F \ne 0$.
In Fig.\ \ref{fig4}(b), we show the calculated $Z_F$ as a function of
the impurity scattering rate.
For $\gamma = 0$, $Z_{F} = 0$ indicating that there is no Fermi
surface, but as scattering is increased $Z_{F}$ also increases
until it saturates at very large $\gamma$ (where our results
should not be trusted because our treatment ignores localization).
Note that $Z_F$ goes to zero
slowly as $\gamma\rightarrow 0$, implying that even a small amount of
impurity scattering results in a fairly pronounced discontinuity in
$n_k$ at $k_F$.  In Fig.\ \ref{fig4}(c), we show the density of states (per
spin)
\begin{equation}
\rho(\omega) = {1\over 2\pi} \int_{-\infty}^\infty {dk\over 2\pi}
A(k,\omega)
\end{equation}
for two different impurity scattering rates.\cite{problem}
For a clean wire, there is a
slow (inverse logarithmic) disappearance in $\rho(\omega)$ as
$\omega\rightarrow 0$, indicating the existence of a gap in the
density of states at the Fermi surface.  For non-zero $\gamma$,
our numerical results indicate that $\rho(\omega)$ initially decays
as $\omega\rightarrow 0$, but then flattens out to a finite value,
indicating the absence of a gap (in accordance with Fermi liquid
theory).  Furthermore, note that in this calculation, we have not
included the effects of scattering in the Green's function in the
self-energy (see subsection \ref{subsec:impure}), which would smear
the density of states further and also remove the gap associated with
the singular behavior of a Luttinger liquid at the Fermi surface.

Fig.\ \ref{fig5}(a) shows the inelastic scattering rates of quasiparticles
in the conduction band $\Gamma(k) = 2\,|{\rm Im}[\Sigma
(k,\omega = \xi_k)]|$
for parameters corresponding to
$a=100\,{\rm\AA}$ and a density of $n = 0.56 \times 10^{6}\,{\rm cm}^{-1}$
in GaAs.
Fig.\ \ref{fig5}(b) shows
the corresponding inelastic mean free path, $l = v(k)/\Gamma(k)$, where
$v$ is the electron velocity.
For $\gamma = 0$, below a threshold wavevector $k_c$, there is {\em no}
electron-electron scattering (within the RPA) because in a strictly
one-dimensional system, conservation of energy and momentum restricts
electron--electron scattering to an exchange of particles, which is not a
randomizing process because electrons are indistinguishable.\cite{leburton}
(Our treatment
ignores multiparticle excitations, which will give rise to a nonzero
scattering rate for $k < k_c$.)
For $k > k_c$, a new scattering channel opens in which electrons genuinely
emit plasmons (as opposed to the virtual plasmon excitations at the Fermi
surface).  The inelastic scattering rate diverges as $\sim (k - k_c)^{-1/2}$
as one approaches $k_c$ from above, due to the divergence in the density of
states available for scattering right at the plasmon emission
threshold, analogous to the divergence in the optical-scattering rate
in one-dimensional systems.\cite{shockley}
In section \ref{sec:sec4}, we discuss this singular behavior in the
scattering rate in greater detail and propose that this, in principle,
can be used to fabricate a novel device.
For $\gamma \ne 0$, the inelastic scattering rate
remains finite because the plasmon line
is broadened.  Furthermore, the breaking of translational invariance
relaxes momentum conservation, permitting inelastic scattering via
single particle excitations for $k < k_c$.

In Fig.\ \ref{fig6}, we show the results of the calculation of the
the bandgap renormalization, which is the sum of ${\rm Re}[\Sigma(k=0,
\omega=\xi_{k=0})]$ of the conduction band electrons and the
valence band holes.
These results should be useful in explaining photoluminescence experiments
in quantum wires, even though our calculation takes into account the screening
effects of electrons only, while a photoexcited
semiconductor in fact contains a finite density of both electrons and holes.
This is justified because we expect the screening effect of
the holes to have a negligible effect on the bandgap renormalization,
since their large mass prevents them from screening effectively.

Table \ref{table2} summarizes the properties of a
quasi one-dimensional quantum wire, for both clean and dirty systems.

\subsection{Impurity effects on the bare Green function}
\label{subsec:impure}

So far, we have ignored the effect of the impurity potential on
the non-interacting Green function $G_0(k,\omega)$;
{\em i.e.}, we have used the $G_0(k,\omega)$ for the
impurity-free case to calculate the many-body properties of impure,
disordered quantum wires.  In fact, the disorder caused by the impurity
potential in one-dimensional systems has a very significant effect,
tending to localize all the eigenstates,\cite{anderson,gof4}
thus significantly changing $G_0(k,\omega)$ from the impurity-free case.
In this subsection, we address the question whether using the
non-disordered $G_0(k,\omega)$ for the disordered case is justified.
We also calculate $G_0(k,\omega)$ for the disordered case for some
simple models.

For a particular impurity configuration, assuming
the exact non-interacting electron eigenstates $\psi_i(x)$ with
eigenenergies $\xi_i$ (with respect to the chemical potential),
the non-interacting Green's function would be diagonal in the basis of
these eigenstates, $G_0(ij,\omega) = \delta_{ij}
\Bigl(\omega-\xi_i + i0^+{\rm sign}(\xi_i)\Bigr)^{-1}$.
Thus, for the non-interacting case, the distribution function at $T=0$
would be a $n_i = \theta(-\xi_{i})$; {\em i.e.}, the momentum index $k$
in the pure case is simply replaced by the indices enumerating the
exact eigenstates.  Hence, our calculation corresponds to the situation
where the $k$ labels in the self-energy correspond not exactly to
the momentum, but to the indices of the exact eigenstates of the
disordered noninteracting system.  Thus, the discontinuity at
the Fermi surface is not a discontinuity in the distribution function
in momentum space, but in the representation of the exact eigenstates
of the disordered system, and our results should be interpreted
in the basis of the disordered eigenstates.
The arguments regarding the suppression of
plasmons in the disordered system should still hold, since the fact that
there are no low energy collective oscillations in disordered systems
is independent of whatever basis is used.

If one wanted to calculate
$\Sigma(k,\omega)$ in the basis of the momentum eigenstates
then the bare Green's function would have to include the
effects of the impurity potential.  In this subsection, we calculate the
distribution function $n_k$, which is the occupation of the
disorder-free momentum eigenstates, for varying amounts of disorder,
using two different methods.  In the first, we use the standard
impurity averaged scheme\cite{agd,doniach} to calculate,
to lowest order in the impurity
scattering potential and in the absence of interactions between the electrons,
the Green's function and the momentum distribution function $n_k.$
The Feynman diagram for this approximation is shown in Fig.\
\ref{fig7}.
We show that, as in higher dimensions, the presence
of impurities broadens $n_k$ because the exact eigenstates are
superpositions of different $k$-states.  Since there are no interactions
between the electrons, the electrons still form a Fermi liquid, in
the sense that there is a sharp discontinuity between the occupation of
states above and below the Fermi energy.
The second method we use to calculate $n_k$ involves putting
in a term representing impurity scattering in the expression
for the exact Green function of the Luttinger liquid.  We show that the
impurity scattering tends to remove the singularities associated with
the non-Fermi liquid nature of the Luttinger liquid.
A comparison of both momentum distribution functions
indicates that for large enough impurity scattering rate,
the noninteracting and interacting distribution functions are
essentially indistinguishable, which implies that any singularities
in $n_k$ associated with the interaction is masked by impurity effects.

For the standard impurity self-energy for impurity scattering
potentials of the form $U_0\delta(r-r_i)$,
\begin{eqnarray}
\Sigma(k,\omega) &=& {N_i |U_0|^2\over 2\pi} \int_{-\infty}^\infty
{dk\over \omega-\xi_k-i0^+}\nonumber\\
&=& -C
\left\{ \begin{array}{ll}
|\omega+\mu|^{-1/2}, &\mbox{if $\omega+\mu < 0$;}\\
i(\omega+\mu)^{-1/2},&\mbox{if $\omega+\mu > 0$;}
\end{array}\right.\label{sigmaimp}
\end{eqnarray}
where $C= N_i U_0^2 \Bigl({m\over 2}\Bigr)^{1\over 2}
= \gamma_{\rm imp}\sqrt{E_F}/2$,
where $\gamma_{\rm imp}$ is the impurity Born approximation scattering
rate for an electron at the Fermi surface.
Defining $\omega' = \omega+\mu$, the spectral function, from
Eq.\ (\ref{spectral})
and (\ref{sigmaimp}), is
\begin{equation}
A(k,\omega=\omega'-\mu) = {2\pi\over 1 + {C\over 2}
|\omega'|^{3/2}}\delta(\omega'-\omega_0) +
{C/\sqrt{\omega'}\over \omega'(\omega'-E_k)^2 + C^2}
\end{equation}
where $\omega_0$ is the solution of
\begin{equation}
\omega_0 - E_k = -{C\over \sqrt{|\omega_0|}}.
\end{equation}
The first term on the right hand side of the above equation is
from a $\delta$-function in the spectral function below the band-edge,
corresponding to a bound state.

The momentum distribution function, from Eq.\ (\ref{nkfroma}), is
\begin{equation}
n_k = {1\over 1 + C|\omega_0|^{3/2}/2} +
\int_0^\mu {d\omega'\over \pi}\; {C \sqrt{\omega'}\over\omega'(\omega'-
E_k)^2 + C^2}.
\end{equation}
The chemical potential $\mu$ is calculated self-consistently so that,
as $\gamma_{\rm imp}$
is changed, the total density $n = \int dk\; n_k/\pi$ is kept constant.
Fig.\ \ref{fig8} shows $\mu$ as a function of $\gamma_{\rm imp}$,
and the corresponding momentum distribution functions.
Note that, by keeping only the first order impurity scattering diagram,
we are assuming weak scattering.  Therefore, the results are unreliable
in the large $\gamma_{\rm imp}$ ({\em i.e.}, large disorder) limit, multiple
scattering events are ignored. For example, if we included all the
diagrams corresponding to repeated scattering off a single impurity,
each such diagram with $n$ impurity lines for a short-ranged interaction
contributes a term $N_i U_0 (i U_0 \sqrt{m/(2\omega+\mu)})^n$, and
therefore the entire series can be summed up to give
\begin{equation}
\Sigma(k,\omega) = {N_i U_0^2\over -i\sqrt{2\omega/m}- U_0}.
\end{equation}
Therefore for large $U_0$, the $\Sigma(k,\omega)$ is modified
significantly from its lowest order form.   There will also be
significant contributions from scattering events off two or more impurities.

We also investigate how disorder affects the distribution function
$n_k$ of the Tomonaga-Luttinger model, where the kinetic energy
of the electrons is given by the completely linear dispersion
$\xi_k = v_F k$.
The exact real-space Green function for the interacting
Luttinger model where the interaction term is
\begin{equation}
\hat H_{\rm e-e} = {1\over 2}\sum_{q,k,k'} g\,{\pi v_F\over 2}\,
\hat a^\dagger_k \hat a^\dagger_{k'} \hat a_{k'-q} \hat a_{k+q}
\end{equation}
and where the interaction $g$
cuts off for momenta larger that $\Lambda$ (which would physically
correspond to the band-width cutoff in a real system, and which
we set equal to $k_F$) can be found
exactly,\cite{larkin,solyom} and it is of the form
\begin{equation}
G(x,t) = G_0(x,t) F(x,t),\label{glut}
\end{equation}
where $F(x,t)$ is dependent on the interaction strength and bandwidth cutoff.

The standard method for introducing impurities into the noninteracting
Green's function is to introduce an electron lifetime $\tau$ to the Green's
function,
\begin{equation}
G(k,\omega) = {1\over \omega-\xi_k + {i\over 2\tau}{\rm
sign}(\omega)}
\end{equation}
which gives a real-space representation of $G(x,t)$ of\cite{agd}
\begin{equation}
G_{\rm imp}(x,t)
= {1\over 2\pi} {1\over x - v_F t + i0^+{\rm sign}(t)}
\exp\Bigl(-{|x|\over 2v_0\tau}\Bigr).
\end{equation}
Therefore, for the Green function in the case where both
interaction and disorder exist, we make the plausible replacement
of $G_0(x,t)$ by $G_{\rm imp}(x,t)$ in Eq.\ (\ref{glut}),
yielding $G(x,t) = G_{\rm imp}(x,t) F(x,t)$.
{}From this, the momentum distribution function is computed to be
\begin{equation}
n_k = {1\over 2} - {1\over 2\pi}\int_{-\infty}^\infty dx
{\sin(kx)\over x} \exp(-{|x|\over 2v_F\tau}) \,\Bigl(1+\Lambda^2
x^2)^{-\alpha},\label{nlut}
\end{equation}
where
\begin{equation}
\alpha = ((1+g)^{1/2}-1)^2/(8(1+g)^{1/2}).\label{31a}
\end{equation}
For $g \ll 1$, $\alpha\approx g^2/32$.

The exact theories assume interactions which are constant
(in momentum space) and short-ranged (in coordinate-space),
which is clearly not the case for the bare Coulomb interaction in a
quasi-one-dimensional quantum wire.
However, in many experimental situations, the quantum wires are surrounded by
mobile charges from objects such as metallic gates or adjacent quantum wires
which would tend to screen the Coulomb interaction between electrons within
the same wire.  This screening would reduce the range of the
interactions, resulting in an effective intrawell interaction $V_{\rm
eff}(q)$ which is finite as $q\rightarrow 0$.
The effective interaction for interaction for a wire screened
by mobile carriers in an adjacent wire, for experimental parameters
given in references \cite{plaut},
is on the order of $g=7.5$, which gives $\alpha=0.15$ (see Appendix
\ref{app:appe}).

In Fig.\ \ref{fig9} we show the the distribution function given by Eq.\
(\ref{nlut}),
for $\alpha=0.15$, and various values of $\gamma$.  For $\gamma = 0$,
the distribution function has an inflection point, $|n_k-1/2| \sim
|k-k_F|^{2\alpha}$ as $k\rightarrow k_F$, which takes the place of the
discontinuity of the non-interacting Fermi liquid.
The introduction of disorder smears this inflection point out, as
it would with a Fermi surface in higher dimensions, and with
increasing disorder it becomes indistinguishable from the $n_k$ shown
in Fig.\ \ref{fig8} for a {\em non-interacting} disordered system.
Thus, impurity effects tend to mask the singular behavior of
a Luttinger liquid, which would tend to complicate any experimental
attempt to measure the the Luttinger liquid singularity in the $n_k$.

\subsection{Discussion of the zero-temperature self-energy}
\label{subsec:discuss}

In this subsection, we compare the relative merits of the
Tomonaga-Luttinger model and our approach, and we discuss
the justification for treating electrons in one-dimensional
semiconductor quantum wires as Fermi liquids.

As mentioned previously, interacting one-dimensional systems
have been shown to be non-Fermi liquids, through the solution
of the Tomonaga-Luttinger model. With the model's assumptions
of infinite density of negative energy electrons
in a completely linear dispersion and short-ranged interactions, all
the vertex corrections in the self-energy are included and the
solution is exact.  On the other hand we assume a finite density of
electrons in a parabolic energy dispersion and we use the {\em actual}
Coulomb interaction\cite{lai} between electrons for a rectangular well,
but we carry out only the leading order self-energy
calculation in the dynamically screened interaction.  Thus, previous
work has concentrated on the exact solution of an artificial
model, whereas our work is an approximate solution of a realistic model.

While the solution of the Tomonaga-Luttinger model is mathematically
exact, it is difficult to see from the exact solution
which physical processes are responsible for the drastic changes
in the low-energy behavior of the system and the disappearance of the Fermi
surface.  Furthermore, since the model linearizes the spectrum about the
Fermi surface, it is not reliable in describing physics at
energies far away from the Fermi energy, when the band parabolicity
becomes important.  While our calculation is only
approximate, we can use it to isolate the processes that have a
a decisive role in determining the low-energy properties of the one-dimensional
interacting electron gas.  We have shown that, physically, the
virtual plasmon emission at low energies is responsible for the
disappearance of the Fermi surface.  In Appendix \ref{app:appf}, we show that
inclusion of leading order vertex corrections in the form used by Mahan and
Sernelius\cite{mahanser} does not change our results at the Fermi
surface.  We can also explore the high energy processes, such as inelastic
scattering of hot electrons, because we have included band parabolicity
and we can calculate the band-gap renormalization effects, because unlike the
Tomonaga-Luttinger model, our model has a band minimum.

We conclude this section by discussing the viability of treating
one-dimensional electron systems as Fermi liquids.
One-dimensional electron systems in solids have {\em theoretically}
strikingly different behavior from their higher dimensional counterparts.
In addition to the Luttinger liquid behavior, there are other effects,
such as the Anderson\cite{anderson,gof4} localization in the
presence of disorder and the Peierls lattice distortion in the presence
of electron--phonon coupling, which theoretically should drive
a one-dimensional electron system away from Fermi liquid behavior.
It would seem to be completely inappropriate to think about one-dimensional
electron systems as Fermi liquids.  We argue, however, that in real
semiconductor quantum wires, various factors may serve to stabilize
Fermi-liquid behavior.

In one dimension, for non-interacting electrons in the
presence of {\em any} disorder (again, invariably present in real wires), all
states are localized and the concept of a one-dimensional electron gas
does not (except in the unrealistic no-impurity idealization) apply.
Therefore, in the absence of electron--electron interactions,
all the quantum wire electronic states are exponentially Anderson-localized.
The situation is again fundamentally different from higher dimensional electron
systems where finite disorder strength is required to strongly localize
the electron states.  We argue, however, that in the state of the art
semiconductor quantum wires the effect of Anderson
localization is negligibly small because typically the localization
lengths are very long (many microns) due to the high quality
of the sample and the modulation doping technique.
Furthermore, the effects of electron--electron interaction on the
localized states of a disordered one-dimensional system are by
no means completely understood, and theories exist which indicate that small
amounts of disorder may not localize an interacting one-dimensional
electron gas.\cite{ren}  In any case, for weakly disordered quantum
wires, the screening effect of the electrons would reduce
the effect of the impurity potential and increase the localization length.
Thus, so long as the localization length is much longer than the length scale
of the experimental probe ({\em e.g.}, in Raman scattering, that would be
the wavelength of the photons), the electrons in the quantum wires may
be considered to be extended for all practical purposes.

The presence of coupling of electrons to the acoustic-phonons
produces, in the ideal zero-temperature situation,
a Peierls distortion,\cite{peierls} in which
the lattice distorts to lower the energy of the electron gas at the
expense of a smaller increase in the lattice potential energy.
The Peierls distorted state has a bandgap at the Fermi wavevector,
and so theoretically quasi-one-dimensional semiconductor quantum wires should
be insulators at $T=0$.  However, the electron-phonon interaction
via the deformation potential coupling is so small that the Peierls
transition temperature is on the order of
$10^{-3}\,{\rm K}$, or much less, depending on the density.\cite{senna}
Furthermore, the transition is suppressed altogether if the impurity
scattering rate is larger than the transition temperature,\cite{suppress}
and therefore in semiconductor quantum wires, since the Peierls
transition temperature is so low, even a small amount of disorder
will remove the transition.

Finally, we have shown that the impurities serve to remove the mechanism
(virtual plasmon emission) responsible for destroying the Fermi surface.
Therefore, as with the Peierls distortion, the presence of impurities
seems to prevent the system from evolving away from Fermi liquid behavior.
Thus, for all practical purposes, real semiconductor quantum wires
can be considered to be normal one-dimensional Fermi liquid systems.

\section{MANY-BODY PROPERTIES OF QUANTUM WIRES\\ AT FINITE TEMPERATURE}
\label{sec:sec3}

In general, many-body calculations
in metals\cite{metal} have used the zero-temperature formalism
because the energy scales intrinsic to the problem
(Fermi energy, plasmon energy, etc) are usually much larger than
the temperature $T$.
In contrast, in semiconductors, especially in artificial structures of
reduced dimensionality, because of the low electron densities and large
lattice dielectric constants involved, the experimental
temperature can be comparable to the intrinsic energy scales of
the electron gas.
For example, for a quantum wire density of $5\times 10^{5}\,{\rm cm}^{-1}$, the
Fermi energies are of order $5\;{\rm meV} \sim 50\;{\rm K}$.
Furthermore, the one-dimensional plasmons are gapless and have a
dispersion that is almost linear, implying that there is a large
density of low-energy collective excitations.
Thus, even at relatively low temperatures, the intrinsic energy scales
of the electrons and of collective excitations
can be comparable to  the temperature.
Therefore, the zero-temperature formalism may not provide an
adequate description of the system, and the finite-temperature
formalism is needed.\cite{dassarma}
In this section, we calculate the finite temperature self-energy, again
using the $GW$ approximation ({\em i.e.,} the
Feynman diagram shown in Fig.\ \ref{fig1}(a)).

In the $T=0$ case, by obeying the Feynman rules,
one obtains an integral expression for $\Sigma(k,\omega)$ in the
$GW$ approximation.  To numerically evaluate $\Sigma(k,\omega)$,
we used the so-called ``line and pole" decomposition,
which is obtained by a contour deformation of the original integral
expression for $\Sigma(k,\omega)$.\cite{mahan,ferrell}  The ``line and pole"
decomposition is more efficient for the purpose of numerical computation
than the original expression.
However, to the best of our knowledge, this ``line and pole"
decomposition has not been generalized, in the published literature,
to finite-temperature self-energies.
In this section, we introduce a general method
for obtaining expressions for the real-frequency
finite-temperature self-energies,
based on analytic continuation of the temperature (Matsubara)
Green's function, which, in the case of the $GW$ approximation, yields the
finite-temperature generalization of the ``line and pole" decomposition.
This method is valid for arbitrary dimensionality.
We present the results of our calculations for a semiconductor quantum
wire using this formalism.

\subsection{Formalism}
\label{subsec:formal}

Matsubara\cite{matsubara} first introduced a many-body
formalism for finite-temperature systems, which is formally
identical to the many-body zero-temperature formalism.
While the finite-temperature formalism is easier to
handle than the zero-temperature formalism in some ways, there
is one added complication.  With the finite-temperature formalism,
one obtains an expression for the electron self-energy $\sigma(i\nu_n)$
that is only valid at discrete points on the complex frequency plane,
$i\nu_n \equiv i(2 n + 1)\pi T$, where $n$ is an integer.
However, the quantity
that is of interest for physical properties of a system is the
self-energy at {\em real} frequencies.
Therefore, the $\sigma(i\nu_n)$ must be {\em analytically continued}
to the complex plane to obtain the self-energy $\Sigma(z)$
that is valid for all complex frequencies,\cite{luttinger}
(for convenience, at this stage, we have suppressed all arguments of
the self-energy except for frequency)
from which the real-frequency (retarded) self-energy,
the quantity relevant to experiments,
can be obtained by setting $z = \omega + i0^+$.
Note that in this section, we use the convention that
functions denoted by upper case characters are analytic in the frequency
variable, while those denoted by lower case characters may be nonanalytic.

In principle, from $\sigma(i\nu_n)$, one can
obtain a formal expression for $\Sigma(z)$ in
terms of integrals over spectral representations.\cite{mahan}
For example, in the $GW$ approximation, the expression for the
real-frequency retarded self-energy is
\begin{equation}
\Sigma({\bf k},\omega) =
\int  {d{\bf q}\over (2\pi)^d}
\int_{-\infty}^\infty {d\omega'\over 2\pi}
{B(q,\omega') \over \omega + \omega' - \xi_{k+q}-i0^+}
\Bigl(n_{\rm B}(\omega') + n_{\rm F}(\xi_{k+q})\Bigr)\label{sfromspec}
\end{equation}
where $d$ is the dimensionality of the system,
$B(q,\omega)$ is the spectral function of the screened Coulomb
interaction, given by
\begin{equation}
B(q,\omega) = 2{\rm Re}[W(q,\omega)] = -i\Bigl(W(q,\omega+i0^+) -
W(q,\omega-i0^+)\Bigr),
\end{equation}
and
\begin{mathletters}
\begin{eqnarray}
n_F(x) &=& {1\over\exp(x/T) + 1}\\
n_B(x) &=& {1\over\exp(x/T) -1 }
\end{eqnarray}
\end{mathletters}
are the bose and fermi functions.
Expressions for $\Sigma(\omega)$ obtained in this manner, however,
involve integration over one or more frequency variables,
which makes them inefficient for use in numerical computations.
On previous occasions various approximations
such as the plasmon-pole approximation
were used to obtain the finite-temperature self-energy.\cite{dassarma}
The plasmon-pole approximation assumes that the spectral function
of the screened Coulomb interaction, $B(q,\omega)$, can be replaced by a
$\delta$-function, whose weight and position are given by sum rules.
The $\delta$-function removes the frequency integration, considerably
simplifying the numerical computation.  Of course, this is an
uncontrolled approximation, and it has been shown that on some occasions
it is quantitatively inaccurate,\cite{inaccurate} and it is desirable
to evaluate the expression Eq.\ (\ref{sfromspec}) efficiently,
without making any approximations.
In this paper, we present a more direct method of analytically continuing
the $\sigma(i\nu_n)$ to $\Sigma(z)$ which yields an exact expression
which is much more amenable to numerical calculation than the
integrals over spectral functions.
We then apply this method to calculate various many-body properties
of electrons in a quantum wire.

This method is based on the principle that the
the analytic continuation of $\sigma(i\nu_n)$ satisfies certain
{\em necessary} and {\em sufficient} conditions, and once
these conditions are satisfied, one is ensured that one has obtained
the unique\cite{baym} analytic continuation, $\Sigma(z)$.
These conditions are as follows.
(1) $\Sigma(z)$ is analytic on the entire complex frequency plane,
with the exception of possible branch cuts on the real axis\cite{luttinger}
(henceforth, when we say a function is ``analytic" it is with the
implicit understanding that it could have branch cuts on the
real axis); (2) $\Sigma(z = i\nu_n) = \sigma(i\nu_n)$ for all $i\nu_n$;
and (3) $\Sigma(z)$ goes to a constant as $|z|\rightarrow\infty$.
The method is based on systematically fulfilling each of the above
conditions one by one, leading us directly from $\sigma(i\nu_n)$ to
its analytic continuation $\Sigma(z)$.

To expound this method, we discuss how it can be applied to a simple example,
that of the self-energy within the $GW$
approximation (see Fig.\ \ref{fig1}(a)) to obtain the finite-temperature
generalization of the ``line and pole" decomposition.
{}From the Feynman rules, the expression for the Matsubara self-energy for
imaginary frequencies is\cite{mahan}
\begin{equation}
\sigma(k,i\nu_n) = -\int {d{\bf q}\over(2\pi)^d}\ \,
T\sum_{i\omega_n} {V(q,i\omega_m)\over\epsilon(q,i\omega_n)}
{1\over i\nu_n + i\omega_n - \xi_{{\bf k}+{\bf q}}},
\end{equation}
where the summation is over the boson frequencies $i\omega_n =
i2\pi n T$ ($n$ are integers).
This expression can be written as a sum of a
frequency-independent (and hence analytic with respect to frequency)
exchange and a frequency-dependent correlation part,\cite{mahan}
\begin{mathletters}\begin{eqnarray}
\sigma({\bf k},i\nu_n)
&=& \sigma_{\rm ex}({\bf k}) + \sigma_{\rm cor}({\bf k},i\nu_n);\\
\sigma_{\rm ex}({\bf k}) &=& - \int {d{\bf q}\over 2\pi}\ \,
V_c({\bf q})\;n_F(\xi_{{\bf k}+{\bf q}}),\\
\sigma_{\rm cor}({\bf k},i\nu_n) &=& -\int {d{\bf q}\over (2\pi)^d}\
h_{{\bf k},{\bf q}}(i\nu_n),\label{sigmacor}
\end{eqnarray}\end{mathletters}
where
\begin{equation}
h_{{\bf k},{\bf q}}(i\nu_n) =
T\sum_{i\omega_n}{w({\bf q},i\omega_n)\over i\nu_n + i\omega_n-\xi_{{\bf
k}+{\bf q}}}.\label{hkq}
\end{equation}
Here,
\begin{equation}
w({\bf q},i\omega_n)= V_{c}({\bf q})
\Bigl({1\over\epsilon({\bf q},i\omega_n)}-1\Bigr)
\end{equation}
is the difference between the screened
and bare Coulomb interactions.
The problem is to analytically continue $\sigma_{\rm cor}({\bf k},i\nu_n)$ to
$\Sigma_{\rm cor}({\bf k},z)$.  The most obvious attempt,
the simple substitution $i\nu_n \rightarrow z$ in Eq.\ (\ref{hkq}),
does not give the desired analytic continuation because there are
poles in the integrand $h_{{\bf k},{\bf q}}(z)$ at $z = \xi_{{\bf k}+{\bf q}} -
i\omega_n$, which translate into branch cuts in $\sigma_{\rm cor}({\bf
k},z)$ at ${\rm Im}[z] = 2\pi n T$ for all $n$.  This clearly violates
the condition (1) stated above.

The $\sigma_{\rm cor}({\bf k},i\nu_n)$ is obtained by an integral over
{\bf q} of an imaginary-frequency-dependent integrand
$h_{{\bf k},{\bf q}}(i\nu_n)$,
as shown in Eq.\ (\ref{sigmacor}).  Thus, if we obtain
analytic continuation $H_{{\bf k},{\bf q}}(z)$ of
$h_{{\bf k},{\bf q}}(i\nu_n)$, then
\begin{equation}
\Sigma_{\rm cor}({\bf k},z) = -\int {d{\bf q}\over (2\pi)^d}
H_{{\bf k},{\bf q}}(z),
\end{equation}
automatically satisfies conditions (1) and (2) above for
the analytic continuation of $\Sigma(k,z)$.\cite{rigour}
Hence, our problem is reduced to constructing function
$H_{{\bf k},{\bf q}}(z)$ such that
(1) it is analytic (in the sense mentioned above, {\em i.e.,} analytic
off the real frequency axis), and
(2) $H_{{\bf k},{\bf q}}(z=i\nu_n) = h_{{\bf k},{\bf q}}(i\nu_n)$.

The outline of the procedure for obtaining the
function $H_{{\bf k},{\bf q}}(z)$ which satisfies conditions \
(1) and (2) is as follows.
First, we write down a function $H^A_{{\bf k},{\bf q}}(z) = h_{{\bf k},{\bf
q}}(z) + \tilde h_{{\bf k},{\bf q}}(z)$,
where $\tilde h_{{\bf k},{\bf q}}(z)$ is chosen so that it
{\em cancels} all the singularities in $h_{{\bf k},{\bf q}}(z)$ on the complex
plane.
Below, we describe a systematic method of obtaining $H^A_{{\bf k},{\bf q}}(z)$
from
$h_{{\bf k},{\bf q}}(z)$.
Thus, $H^A_{{\bf k},{\bf q}}(z)$ is {\em analytic}, fulfilling condition (2).
However, $\tilde h_{{\bf k},{\bf q}}(z)$ is in general nonzero at $z=i\nu_n$,
and hence
$H^{\scriptscriptstyle{\rm A}}_{{\bf k},{\bf q}}(i\nu_n)
\ne h_{{\bf k},{\bf q}}(i\nu_n)$, violating condition (2).
The second step is therefore to add
an additional analytic term $H'_{{\bf k},{\bf q}}(z)$ which {\em cancels}
$\tilde h_{{\bf k},{\bf q}}(z)$ at all $z = i\nu_n$.  Since the function
$H^A_{{\bf k},{\bf q}}(z) + H'_{{\bf k},{\bf q}}(z)$
is analytic {\em and} equals $h_{{\bf k},{\bf q}}(z)$
for all $z=i\nu_n$, fulfilling both conditions (1) and (2),
it is the desired analytic continuation $H_{{\bf k},{\bf q}}(z)$.
Condition (3) can be checked in the end; in the case of the $GW$
approximation (and in other cases we have studied) it is satisfied.

In the case of the $GW$ self-energy, the expression obtained from
the Feynman rules is
\begin{equation}
h_{{\bf k},{\bf q}}(z) =
T\sum_{i\omega_n}{w({\bf q},i\omega_n)\over z + i\omega_n-\xi_{{\bf k}+{\bf
q}}}.\label{defhkqz}
\end{equation}
One can construct an analytic function which contains $h_{{\bf k},{\bf q}}(z)$
as one
of its terms by writing a contour integral
\begin{equation}
H^A_{{\bf k},{\bf q}}(z)
= \int_{\cal C} {d\omega\over 2\pi i} {w({\bf q},\omega)
n_B(\omega) \over z + \omega - \xi_{{\bf k}+{\bf q}}},\label{contour}
\end{equation}
where the contour of integration ${\cal C}$ is shown in Fig.\ \ref{fig10}.
$H^A_{{\bf k},{\bf q}}(z)$ is clearly analytic (off the real axis)
in the variable $z$,
since the function $(z + \omega - \xi_{{\bf k}+{\bf q}})^{-1}$ has no poles off
the
real axis.
By the residue theorem, the integral Eq.\ (\ref{defhkqz}) can be evaluated in
terms of the residues of the poles contained within the contour ${\cal C}$.
This gives
$H^A_{{\bf k},{\bf q}}(z) = h_{{\bf k},{\bf q}}(z) +
\tilde h_{{\bf k},{\bf q}}(z)$.
The term $h_{{\bf k},{\bf q}}(z)$, comes from the poles of $n_B(\omega)$,
as can be seen more clearly by writing
$n_B(\omega)$ as the sum of its poles,
\begin{equation}
n_B(\zeta) = -{1\over 2} + T\sum_{i\omega_n}  {1\over\zeta
- i\omega_n}.\label{boseexp}
\end{equation}
In addition to the poles of the bose function, there is also a pole due to the
denominator of Eq.\ (\ref{defhkqz}) which gives the second term
\begin{equation}
\tilde h_{{\bf k},{\bf q}}(z) = w({\bf q},\xi_{{\bf k}+{\bf q}}-z)\;
n_B(\xi_{{\bf k}+{\bf q}} - z).
\end{equation}
(Note that $w({\bf q},\omega)$ is analytic everywhere
except for a branch cut on the real axis, and so it does not
have poles which contribute to $H^A_{{\bf k},{\bf q}}(z)$.)
Despite being the sum of two nonanalytic functions,
$H^A_{{\bf k},{\bf q}}(z)$ is analytic because the
singularities present in both
the functions precisely cancel each other.
This can be seen clearly if we use Eq.\ (\ref{boseexp}) to expand the bose
function of $\tilde h_{{\bf k},{\bf q}}(z)$, yielding
\begin{eqnarray}
H^A_{{\bf k},{\bf q}}(z) &=&
h_{{\bf k},{\bf q}}(z) + \tilde h_{{\bf k},{\bf q}}(z)\nonumber\\
&=& -{1\over 2} w({\bf q},\xi_{{\bf k}+{\bf q}}-z) +
T \sum_{i\omega_n}
{w({\bf q},i\omega_n) - w({\bf q},\xi_{{\bf k}+{\bf q}}-z)\over
z+i\omega_n - \xi_{{\bf k}+{\bf q}}},\label{defhakqz}
\end{eqnarray}
which explicitly shows that both the denominator and numerator
vanish at $z =\xi_{{\bf k} +{\bf q}} - i\omega_n$.

However, because $\tilde h_{{\bf k},{\bf q}}(i\nu_n) \ne 0$,
$H^{\rm A}_{{\bf k}+{\bf q}}(i\nu_n) \ne h_{{\bf k},{\bf q}}(i\nu_n)$
and hence $H^A_{{\bf k},{\bf q}}(z)$ fails to fulfil condition (2).
This can be remedied by adding an analytic function which cancels
$\tilde h_{{\bf k},{\bf q}}(z)$ at all $z = i\nu_n$.
Noting that $i\nu_n = i(2n + 1)\pi T$, for all integers $n$,
\begin{eqnarray}
n_B(\xi_{{\bf k}+{\bf q}}-i\nu_n)
&=& \Bigl[\exp(\xi_{{\bf k}+{\bf q}}/T)\exp(-(2n+1)\pi)
-1\Bigr]^{-1}\nonumber \\
&=& -\Bigr[\exp(\xi_{{\bf k}+{\bf q}}/T) +1\Bigr]^{-1}
= -n_F(\xi_{{\bf k}+{\bf q}})\label{bosfer}
\end{eqnarray}
and thus
$\tilde h_{{\bf k},{\bf q}}(i\nu_n) = -w({\bf q},\xi_{{\bf k}+{\bf
q}}-i\nu_n)\;
n_F(\xi_{{\bf k}+{\bf q}}),$
implying that the analytic term needed to cancel
$\tilde h_{{\bf k},{\bf q}}(i\nu_n)$
is
\begin{equation}
H'_{{\bf k},{\bf q}}(z) = w({\bf q},\xi_{{\bf k}+{\bf q}}-z)
n_F(\xi_{{\bf k}+{\bf q}}).\label{defhpkqz}
\end{equation}
Hence, $H_{{\bf k},{\bf q}}(z)  =
H^A_{{\bf k},{\bf q}}(z) + H'_{{\bf k},{\bf q}}(z)$ is given by Eqs.\
(\ref{defhakqz}) and (\ref{defhpkqz}),
and the correlation self-energy in the $GW$ approximation is
\begin{eqnarray}
\Sigma_{\rm cor}({\bf k},z)
&=& -\int {d{\bf q}\over(2\pi)^d}\ \Bigl[{H^A_{{\bf k},{\bf q}}(z) +
H'_{{\bf k},{\bf q}}(z)}\Bigr]\nonumber\\
&=& -\int{d{\bf q}\over(2\pi)^d} T\sum_{i\omega_n}
{w({\bf q},i\omega_n)\over z + i\omega_n-\xi_{{\bf k}+{\bf q}}}\nonumber\\
&& -\int{d{\bf q}\over(2\pi)^d}\
w({\bf q},\xi_{{\bf k}+{\bf q}}-z)\Bigl(n_B(\xi_{{\bf k}+{\bf q}} - z)
+ n_F(\xi_{{\bf k}+{\bf q}})\Bigr).\label{ftlap}
\end{eqnarray}

The retarded self-energy is obtained by setting
$z \rightarrow \omega + i 0^+$.  The first and second terms
on the right hand side of the last equality
of Eq.\ (\ref{ftlap}) are, respectively,
the finite-temperature generalization of the so-called ``line" and ``pole"
components of the $GW$ approximation of the $T=0$
correlation self-energy.\cite{ferrell}
As in the zero-temperature case, the line contribution is completely
real because $w({\bf q},-i\omega_n)=w^*({\bf q},i\omega_n)$
and hence the total contribution to the
imaginary part of $\Sigma({\bf k},\omega)$ comes from the pole part.

As in the $T=0$ case, in the $GW$ approximation
the ``on-shell" imaginary part of the
self-energy, $|{\rm Im}[\Sigma({\bf k},\omega = \xi_{\bf k})]|$,
is half the sum of the Born-approximation electron and
hole scattering rates.
The Born-approximation rate for
a particle to be scattered with change in momentum ${\bf q}$ and energy
$E$ in the particle is\cite{pines1}
\begin{equation}
P({\bf q},E) = -2 V_c({\bf q}) {\rm Im}[\epsilon^{-1}({\bf q},E)]n_B(E)
\end{equation}
Using the identity
\begin{equation}
n_B(\xi_{{\bf k}+{\bf q}}-\omega) + n_F(\xi_{{\bf k}+{\bf q}})
\equiv
n_B(\xi_{{\bf k}+{\bf q}}-\omega)\Bigl[1-n_F(\xi_{{\bf k}+{\bf q}})\Bigr]
-n_B(\omega-\xi_{{\bf k}+{\bf q}})n_F(\xi_{{\bf k}+{\bf q}}),
\end{equation}
we can write $2\;|{\rm Im}[\Sigma({\bf k},\xi_{\bf k})]|
= \gamma_{\rm e}({\bf k}) + \gamma_{\rm h}({\bf k})$
where
\begin{mathletters}\begin{eqnarray}
\gamma_{\rm e}({\bf k}) &=&
-\int {d{\bf q}\over(2\pi)^d}\
V_c({\bf q})\ {\rm Im}[\epsilon^{-1}({\bf q},
\xi_{{\bf k}+{\bf q}}-\xi_{{\bf k}})]\,
n_B(\xi_{{\bf k}+{\bf q}} -\xi_{\bf k})
\Bigl[1-n_F(\xi_{{\bf k}+{\bf q}})\Bigr],\nonumber\\
&=&\int {d{\bf q}\over(2\pi)^d} P({\bf q},\xi_{{\bf k}+{\bf q}})
\Bigl[1-n_F(\xi_{{\bf k}+{\bf q}})\Bigr],\\
\gamma_{\rm h}({\bf k}) &=&
\int{d{\bf q}\over(2\pi)^d}\ V_c({\bf q})
{\rm Im}[\epsilon^{-1}({\bf q},\xi_{{\bf k}+{\bf q}}-\xi_{{\bf k}})]
n_B(\xi_{{\bf k}}-\xi_{{\bf k}+{\bf q}}) n_F(\xi_{{\bf k}+{\bf q}})\nonumber\\
&=&\int{d{\bf q}\over(2\pi)^d}\ \
P({\bf q},\xi_{\bf k}-\xi_{{\bf k}+{\bf q}})n_F(\xi_{{\bf k}+{\bf q}})
\label{defgamma}
\end{eqnarray}\end{mathletters}
are the total Born approximation electron and hole scattering rates,
respectively.\cite{subtle}
The factors of $1-n_F(\xi_{{\bf k}+{\bf q}})$ and
$n_F(\xi_{{\bf k}+{\bf q}})$ that multiply the Born approximation scattering
rates
in Eq.\ (\ref{defgamma}) are due to the Pauli exclusion principle.
An electron (hole) can scatter to a final state ${\bf k}+{\bf q}$ only if it
is unoccupied (occupied), and the factor $1-n_F(\xi_{{\bf k}+{\bf q}})$
($n_F(\xi_{{\bf k}+{\bf q}})$) is the probability that the final state is
unoccupied (occupied) by an electron from the system.

The method outlined above which we have applied to the $GW$ approximation
can also be applied to higher-order diagrams, although the procedure
quickly becomes considerably more complicated.
In a higher-order self-energy with an external Matsubara Fermi
frequencies $i\nu_m$, the Feynman rules result in an expression
with a summation of a function $f(i\nu_m,\omega_{n_1}, ... ,\omega_{n_N})$,
over several boson frequencies $\omega_{n_1},...,\omega_{n_N}$,
\begin{equation}
h(i\nu_m) = \sum_{i\omega_{n_1}}... \sum_{i\omega_{n_N}}
f(i\nu_m,\omega_{n_1}, ... ,\omega_{n_N}) \label{defhinum}
\end{equation}
In $f(i\nu_m,\omega_{n_1}, ... ,\omega_{n_N})$,
there are several bare Green's function
which contain both $i\nu_m$ and one (or more)
internal Bose frequency $i\omega_{n_i}$; {\em e.g.} terms of the form
$(i\omega_{n_i} + i\nu_m - \xi_{{\bf k}+{\bf q}})^{-1}$.
(One can always write down
the self-energy in a form such that the external frequency occurs
only in the bare Green's functions, and not in the screened
Coulomb interactions.)
To obtain the {\em analytic} function $H^{\rm A}(z)$ which contains
$h(z)$ as one of its term, one writes down an integral
\begin{equation}
H^{A}(z) = \int_{{\cal C}_1} d\omega_1 ...
\int_{{\cal C}_N} d\omega_N\ n_B(\omega_1) ...
n_B(\omega_N) f(z,\omega_1,...,\omega_N),
\end{equation}
where every summation over $i\omega_{n_i}$,
has been replaced replaced by an integral over $\omega_i$,
$i\nu_m$ is replaced by $z$
and $f$ is multiplied by the a product of bose factors.
Since $z$ only appears in the integrand in the form of
factors $(\omega + z - \xi_{{\bf k}+{\bf q}})^{-1}$,
$H^{\rm A}(z)$ is analytic off the real axis.
By using the residue theorem,
one can evaluate the integral in terms the poles of the integrand.
The $n_B(\omega_i)$ terms give summations over Bose functions,
which correspond exactly to the original expression, Eq.\
(\ref{defhinum}), with $i\nu_m$ replaced by $z$.
There are also other terms generated which serve to cancel the
singularities of Eq.\ (\ref{defhinum}) on the complex plane.
As in the example cited above for the $GW$ approximation, this function
fails condition (2), and therefore analytic functions must be found that
cancel the unwanted terms on the points $z=i\nu_m$.  Generally, this can
be done by inspection of the resulting term,
and by use of Eq.\ (\ref{bosfer}).  We have used this
method to obtain an expression for the self-energy which is
second-order in the screened interaction, which we give in
Appendix \ref{app:appg}.

Alternatively, one can also derive the finite-temperature ``line and
pole" expression of self-energy from the spectral representation
of the $GW$ approximation of the self-energy,
given by Eq.\ (\ref{sfromspec}).  The integral
over the spectral function for the Coulomb interaction can
be rewritten as a contour integration over the contour indicated in
Fig.\ \ref{fig10}, with the spectral function in the integrand replaced by
$i^{-1}w({\bf q},\omega)$. Then, by the residue theorem,
one obtains precisely the expression obtained above in Eq.\ (\ref{ftlap}).
This procedure also works for higher order diagrams.
One first obtains the expression for the self-energy in terms of
integrals over spectral functions, and then by writing these
integration of spectral functions as contours of the form shown in
Fig.\ \ref{fig10} and evaluating the contributions due to the poles within the
contour, one obtains an expression involving sums over imaginary frequencies.

The main advantage of the expressions generated by the method outlined
above is that the frequency integrals in the spectral representation method
are generally replaced by imaginary frequency summations, which are
numerically more efficient to perform, especially at high temperatures.
Unfortunately, with the higher order diagrams, there are terms that
still involve frequency integrations, but the number of these frequency
integrations is reduced from the spectral representation expression
of the self-energy.

\subsection{Results for quantum wire}
\label{subsec:results}

We applied the formalism developed above to electrons in a semiconductor
quantum wire.  As mentioned previously,
the densities of current quantum wires are necessarily low,
and the Fermi energies are on the order of $5\;{\rm meV} \sim 50\;{\rm K}$.
Thus, even small temperatures may affect their many-body
properties significantly.
We applied Eq.\ (\ref{ftlap}) to the calculation of the self-energy, spectral
function and band-gap renormalization of electrons in a one-dimensional quantum
wire in the extreme quantum limit ({\em i.e.}, assuming that
the electrons only occupy the lowest energy subband).

We used the RPA form for the dielectric function $\epsilon(q,z)
=1-V_c(q)\Pi_0(q,z)$.
Using the familiar expression for $\Pi_0(q,z)$,
given by Lindhard,\cite{lindhard,mahan}
to numerically compute the real part of
the polarizability would have been complicated by a principal part
divergence in the integrand. Therefore, instead, we used
\begin{equation}
\Pi_0(q,z; T,\mu) =
\left\{\begin{array}{lr}
\displaystyle\int_{e^{-(\mu/T)}}^1 {dx\over (x+1)^2}
\Pi_0(q,z; T=0, \mu + T \ln(x))&\nonumber\\
\displaystyle\qquad\qquad
+ \int_0^1 {dx\over (x+1)^2} \Pi_0(q,z; T=0, \mu - T\ln(x)),
&\mbox{if $\mu > 0$};\nonumber\\ & \\
\displaystyle\int_0^{e^{(\mu/T)}} {dx\over (x+1)^2}
\Pi_0(q,z; T=0, \mu - T\ln(x)),
&\mbox{if $\mu <0$;}
\end{array}\right.
\end{equation}
which is a computationally efficient
version of the expression given by Maldague\cite{maldague}
for the finite-temperature RPA polarizability.
In performing the summations over imaginary frequencies of the ``line"
part of $\Sigma(k,\omega)$, we used the fact that the
asymptotic behavior of the components go as $n^{-5/2}$ as
$n\rightarrow \infty$ (the asymptotic form for these components
for an infinitesimally thin $d=2$ electron gas is $n^{-2}$
and for $d=3$ is $n^{-5/2}$),
which allowed us to use a slightly modified form of the Richardson
extrapolation procedure\cite{bender} to obtain rapid convergence
of the sum.  Also, since there are mutually canceling
principal part divergences
(in the $q$-integration) in the $n=0$ term of the ``line" part
and the real part of the ``pole" term, it
is advantageous to numerically evaluate both these terms together.

We calculated the real and imaginary parts of the self-energy
and the spectral function of a quantum wire
as a function of frequency, for several temperatures.
The density was kept constant by adjusting the non-interacting electron
gas chemical potential $\mu_0(T)$
so that the integral $\int dk\  n_{\rm F}(E_k-\mu_0(T))/\pi$
was kept constant.
We show our results in Fig.\ \ref{fig11}.
The discontinuities and kinks in ${\rm Im}[\Sigma(\omega)]$ at $T=0$,
which arise from virtual plasmon and single-particle emission thresholds,
broaden with increasing temperature because the plasmon peaks
are broadened by Landau damping, and the boundaries of the
single-particle continuum are thermally smeared.
At non-zero temperatures, a logarithmic divergence develops
in the imaginary part of the self-energy at $\omega = \xi_k [\equiv E_k -
\mu_0(T)]$, and is accompanied by a discontinuity in
${\rm  Re}[\Sigma(k,\xi_k)]$, since the real and imaginary parts of
$\Sigma(k,\omega)$ are related by the Kramers-Kronig relations.
This divergence is due to the non-integrable $|q|^{-1}$ singularity
as $q\rightarrow 0$ in the integrand Eq.\ (\ref{ftlap}), which arises
from the product of the bose factor, which goes as $q^{-1}$,
and the ${\rm Im}[\epsilon^{-1}(q,\xi_{k+q}-\xi_k)]$,
which goes as ${\rm sign}(q)$.
Physically, this singularity corresponds to the singular
divergence in the Born-approximation electron--electron scattering
at small momentum transfer in $d=1$, and is
unique to one-dimensional systems, since in higher dimensions,
the singularity in the integral is removed by the phase-space factor
of $q^{d-1}$.  It is not obvious whether this singularity exists
in the exact $\Sigma(k,\omega)$ or is an artifact of the approximations
we have used.

In order to calculate the spectral function, one needs to know
the finite temperature chemical potential for the {\em interacting} system.
At $T=0$, the chemical potential $\mu$ of the interacting system
is simply given by ${\rm Re}[\Sigma(k_F,\omega=0)]$ (as is required to
form a discontinuity or a singularity at the Fermi surface).
Unfortunately, at finite temperatures, there is no such obvious prescription to
find $\mu(T)$ for the interacting system.  In principle,
one could search for the $\mu(T)$ which allows the sum rule
Eq.\ (\ref{asumrule}) to be satisfied for all $k$
(we expect Eq.\ (\ref{asumrule}) to be
satisfied because we are using consistent particle-conserving
approximations\cite{conserve}).  However, in practice,
the integrated spectral weights are not particularly sensitive to
changes in the chemical potential and we were not able to
determine $\mu$ accurately in the attempts we made.
By the same token, the spectral functions themselves are not very
sensitive to the choice of chemical potential.
We used the $T=0$ value of the chemical potential of the interacting
system, with a temperature correction based on
the temperature dependence of $\mu_0(T)$ for the non-interacting system.
As expected, the spectral functions broaden with increasing
temperature, corresponding to an increase in the quasiparticle decay
rates caused by the presence of thermal excitations.
The dips at $\omega=\xi_k$, which go as $\sim|\ln(\omega-\xi_k)|^{-1}$,
which are brought about by the logarithmic divergences of
${\rm Im}[\Sigma(k,\omega)]$.

In Fig.\ \ref{fig12}, we show the electron and hole
band-gap renormalization due to
the presence of conduction electrons alone.  We note again that holes
are not expected to change the results significantly because holes
screen weakly.  Due to the discontinuity in ${\rm Re}[\Sigma(k,\xi_k)]$,
we take ${1\over 2}\ {\rm Re}[\Sigma(k,\xi_k + 0^+) + \Sigma(k,\xi_k - 0^+)]$
at $k=0$ to be the band-gap renormalization.  We find that for
very low densities, where the Fermi temperature is low, the
band-gap renormalization can change by approximately an order of
magnitude when the temperature increases from $T=0\,{\rm K}$ to
$T=300\,{\rm K}$.  Such changes should be measurable in
photoluminescence experiments.

\section{DEVICE POSSIBILITIES FOR QUANTUM WIRES}
\label{sec:sec4}

Part of the reason why so much excitement has been generated by quantum
wires is the possibility that they might have novel device applications.
In the early eighties, Sakaki\cite{sakaki} argued that electrons in
quantum wires could have extremely large mobilities because
elastic collisions with small momentum transfer are suppressed in
one-dimensional systems, which would make them very attractive for
use in semiconductor devices.
More recently, electronic devices based on the quantum nature of
electrons\cite{sol} and on the single-particle Coulomb charging energies
in mesoscopic systems\cite{turnstile} have been proposed.\cite{reviews}
In addition, various proposals for uses of quantum wires as lasers and
optical switches have been advanced.\cite{leburton1}
In this section, we propose another interesting device principle of a
one-dimensional quantum wire, based on the the many-body properties
which are peculiar to quasi-one-dimensional systems.

The device principle we propose is based on the many-body properties
an interacting one-dimensional Fermi system.
We show that it may be possible to obtain a device with
large and sudden onset of negative differential resistance (NDR)
({\em i.e.}, $dI/dV < 0$).
This sudden onset of NDR could be exploited to produce a transistor,
while the NDR itself suggests that this device might be used as an
oscillator ({\em e.g.}, in analogy with the Gunn oscillator\cite{gunn} or,
more recently, the resonant tunneling diode\cite{sollner}).
In the proposed device principle, the predicted NDR is associated with a
{\em sharp} change in the inelastic mean free path, due to electron--electron
interactions (or more specifically, electron--{\em plasmon}
scattering) of the injected electrons
at a specific injection energy --- in the ideal system at $T=0$,
the mean free path changes from being {\em infinite} below the threshold
energy to being {\em zero} above it.

The device principle which we propose may be experimentally
observed in the quasi-one-dimensional version of the
tunneling hot electron transistor amplifier (THETA), shown
schematically in Fig.\ \ref{fig13}(a), which has
been fabricated successfully in three and two dimensions.\cite{heiblum}
As in previous sections, we assume that the quasi-one-dimensional device
is in the extreme quantum limit; {\em i.e.}, that all the electrons are
in the lowest energy subband in the device.
Electrons are injected from an emitter at energies above the Fermi
energy $E_{\rm F}$ into a base region which contains
(either through doping or electrostatic confinement) a one-dimensional
electron gas, and the injected electrons that travel through the base region
enter the collector on the opposite side of the base.
The fraction of electrons $\alpha$ that reach the collector goes as
$\alpha \sim e^{-d/l}$, where $d$ is the width of the base region and
$l$ is the electron mean free path.
The $l$ is equal to $v_k \Gamma(k)$,
where $v_k$ is the electron velocity and $\Gamma(k)$ is the
inelastic scattering rate, which is a strong function of the injection energy
of the electrons and the electron density in the base region.  Thus,
by varying the scattering rate by changing the injection energy,
one can achieve a significant change
in the electron mean free path and hence the emitter--collector current.

In two and three dimensions, the main scattering mechanism for
these electrons in the THETA devices are the coupled plasmon--optic-phonon
modes.\cite{dassarma1}  However, in the semiconductor quantum wires
in the extreme quantum-limit that are currently being fabricated,
the densities of the electrons in the base are so low that all the
energy scales associated with the electron gas and operation of the
device ($E_{\rm F}$, plasmon energy and electron injection energy) are much
smaller than the optic-phonon energy, and
therefore the optic-phonons play a negligible r\^ ole.
Acoustic phonons couple very weakly to electrons in III-V
semiconductors,\cite{seeger}
and the associated scattering rates are on the order of
$10^{10}\,{\rm s}^{-1}$ and can be ignored when compared to the
scattering mechanism discussed in this paper.
We assume, for the purpose of this paper,
that impurity scattering in the wires is
negligible, which is not unreasonable given the excellent and
continually improving techniques for fabricating these mesoscopic
systems.   This last assumption is equivalent to assuming that the {\em
elastic} mean free paths are much longer than the inelastic mean free
path to be calculated in this paper --- given that our calculated
inelastic mean free paths are generally
a few thousand \AA\ or less, and in good quality
quantum wires, elastic mean free paths are many microns, the neglect of
impurity scattering is a good approximation for our purposes.
Furthermore, the restriction of the scattering phase-space in
one-dimension further reduces the scattering rates of both impurity
and acoustic phonons.\cite{sakaki}
Thus the main scattering mechanism for an injected electron is the
interaction with the electron gas in the base.

We use the Born approximation in our calculation of the
scattering rate of the injected electron with the electron gas.
The Born approximation calculation of the scattering rate is equivalent
to the calculation of ${\rm Im}[\Sigma]$ in the $GW$ approximation
described in earlier sections of this paper.
This approximation includes single-particle scattering and scattering with
collective modes of the plasma, but ignores multi-particle excitations,
which are expected to be smaller because they are higher order in the
screened interaction.
However, as we have mentioned in the previous section, in one dimension
and at finite temperature, a straightforward integration of the
Born approximation scattering probability over all wavevectors
yields a total Born scattering rate that is {\em infinite}, due to
the presence of a non-integrable divergence in the number of small wavevector
scattering events.
These small-wavevector scattering events are associated with
single-particle scatterings (in which the injected particle scatters
with an individual electron in the electron gas in the base).
The problem is that the Born approximation treats the injected particle
as an {\em external distinguishable particle}, and
gives the rate at which this particle is scattered out of its initial
state. However, in actual fact, the injected electron is
indistinguishable from the other electrons in the base, which is
very significant in one-dimensional systems, because a single-particle
electron--electron
scattering event involves an exchange of electrons, implying that
the final state after the scattering is exactly the same as the initial state.
Therefore, a scattering rate associated with single-particle
electron--electron collisions in one-dimension is spurious because
these ``collisions" do not change the state of the system.
It is the transfer of momentum of the injected the particle
to the {\em plasmons} of the system that degrades the current
that flows from the emitter to the collector.
The plasmons are collective modes associated with the electrons in the
base and cannot exist outside the base region, and hence cannot
carry current into the collector.
Therefore, to determine the mean free path of the nonequilibrium
distribution of electrons injected into the base, we should only
include the scattering rate of the injected electron with plasmons.
Unfortunately, within the Born approximation at finite temperatures,
there is no way of separating the spurious single-particle scattering
from the plasmon scattering, since the sharp boundaries
and line-widths of the single-particle continuum and plasmon dispersion
which exist at $T=0$ both become thermally smeared and
merge into each other as the temperature increases.\cite{smear}

Our solution to this problem is to calculate the momentum scattering
rate, which is the integral of the Born approximation
probabilities\cite{pines} weighted by the loss of momentum,
\begin{equation}
\Gamma_{\rm t}(k) = 2 \int_{-\infty}^\infty dq\;
{q\over k}\; V_c(q)\; {\rm Im}
\Bigl[\epsilon^{-1}\bigl(q,\omega_k(q)\bigr)\Bigr]
\;n_B(\omega_k(q))[1-n_F(\xi_{k+q})],\label{momscatrate}
\end{equation}
where $k$ is the momentum of the injected electron,
$\omega_k(q) = \xi_k -\xi_{k-q}$ is the energy lost by the electron when
is loses momentum $q$, and $\epsilon(q,\omega)$ is the dielectric
function calculated within the RPA.
$\Gamma_{\rm t}(k)$
is generally the quantity that is relevant for transport, since it
measures the degradation of the electron current.
The form Eq.\ (\ref{momscatrate}) negates the effect of the spurious small
$q$ single-particle
scattering events by giving them small weights and by
canceling momentum loss against momentum gain.  The main contribution
to $\Gamma_{\rm t}(k)$ should come from coherent peaks associated with
the scattering of plasmons.  In Fig.\ \ref{fig14},
we show the results of our calculation of
$\Gamma_{\rm t}(k)$, and the corresponding inelastic mean free path,
$l_k = v_k/\Gamma_{\rm t}(k)$, for various temperatures.

We concentrate on the zero-temperature $\Gamma_{\rm t}(k)$, the
simplest case.
At $T = 0$, there are no thermally excited particles that
contribute to the spurious single-particle scattering rate, and
hence the only contribution to $\Gamma_{\rm t}(k)$
comes from the plasmon emission.
However, not all injected electrons can emit plasmons.
Because the plasmon dispersion in quasi-one-dimensional systems
for small $q$ goes as
$\omega(q) \sim q |\log(qa)|^{1/2}$, where $a$ is the width of the wire,
only injected electrons with large enough kinetic energies
can emit plasmons (see Fig.\ \ref{fig13}(b)).
For a given density $n$, there is therefore
a threshold wavevector $k_c(n)$ below which no plasmon emission can take place.
Within the approximations we have used and at $T=0$,
as $k$ is increased through $k_c$,
the scattering rate jumps from {\em zero} to {\em infinity}
(equivalently, the mean free path falls from {\em infinity} to {\em
zero}). The divergence in $\Gamma_{\rm t}(k)$ as $k\rightarrow k_c^+$,
is $(k - k_c)^{-1/2}$.
Below, we derive this form of the divergence,
and we show that under very special circumstances,
this divergence may be stronger.

At $T=0$, the electrons can only scatter by plasmon emission
if momentum and energy conservation of the system is obeyed
(and the final state of the electron $k-q$ is initially unoccupied).
One can graphically determine if it is possible to emit plasmons
by plotting, on the same graph,
the plasmon dispersion curve, $\omega_{\rm p}(q)$
[given by $\epsilon(q,\omega_{\rm p}(q)) = 0$], and the energy-loss
vs.\ momentum-loss curve for an electron with initial momentum $k$,
$\omega_k(q) = \xi_k - \xi_{k-q}$.
Energy and momentum conservation of the system is obeyed
and plasmon emission is allowed only when the curves
$\omega_{\rm p}(q)$ and $\omega_k(q)$ intersect.
The wavevectors $q^*$ at which $\omega_{\rm p}(q)$ and
$\omega_k(q)$ intersect correspond to wavevectors of plasmons
which the electrons are permitted to emit.
The scattering rate at $T=0$ due to plasmon emission
is given by
\begin{equation}
\Gamma_{\rm p}(k) =  \sum_{q^*} V_c(q^*(k))
\Bigl|{\partial\epsilon\Bigl(q,\omega_k(q)\Bigr)\over
\partial q}\Bigr|_{q=q^*(k)}^{-1},\label{gammapk}
\end{equation}
where the summation is over all intersections $q^*$.
We are interested in finding the $T=0$ behavior of $\Gamma_{\rm p}(k)$ near
$k=k_c$, the minimum injected electron momentum at which the intersection
between $\omega_k(q)$ and $\omega_{\rm p}(q)$ occurs.

Defining $q_c$ and $\omega_c$ as the wavevector and energy of the
at which the intersection of $\omega_{\rm p}(q)$ and
$\omega_{k}(q)$ occurs for $k=k_c$, the following
relations hold:
\begin{mathletters}\begin{eqnarray}
\omega_c &=& {1\over m}\Bigl(k_c q_c - {1\over 2}q_c^2\Bigr),\\
\epsilon(q_c,\omega_c) &=& 0,\\
{\partial \omega_{k_c}(q)\over \partial q}\Bigr|_{q=q_c} &=&
{\partial \omega_{\rm p}(q)\over\partial q}\Bigr|_{q=q_c}.\label{eqnkc}
\end{eqnarray}\end{mathletters}
The first two equations are immediate consequences of the above definitions.
The last equality, Eq.\ (\ref{eqnkc}), results from the fact that, at
$k=k_c$, when $\omega_k(q)$ just impinges on the $\omega_{\rm p}(q)$,
the slopes of the two curves are equal (as can be seen by an
inspection of Fig.\ \ref{fig13}(b)).

Let $\omega_1$, $q_1$ and $k_1$ be the deviations away from
$\omega_c$, $q_c$ and $k_c$,
\begin{eqnarray}
\omega &=& \omega_c+\omega_1,\nonumber\\
q &=& q_c+q_1,\\
k &=& k_c+k_1.\nonumber
\end{eqnarray}
In terms of $q_1$ and $k_1$, the
energy vs.\ momentum loss curve is
\begin{equation}
\omega_{k}(q)=\omega_c + {1\over m}
\Bigl(q_0 k_1 +(k_0-q_0)q_1 + k_1 q_1 - {1\over 2} q_1^2\Bigr).
\label{eqnwkq}
\end{equation}
The dielectric function can be expanded
around $q_c$ and $\omega_c$ in the form
\begin{equation}
\epsilon(q,\omega) \approx -a q_1 + b \omega_1
+ O(q_1^2,\omega_1^2,q_1\omega_1),\label{eqneps}
\end{equation}
where $a,b >0$ (this inequality can be deduced from the asymptotic
behavior of $\epsilon(q,\omega)$).
Eq.\ (\ref{eqneps}) implies that the slope is at
$\omega_{\rm p}(q_c)$ is $a/b$
and hence, by Eq.\ (\ref{eqnkc}),
\begin{equation}
{k_0-q_0\over m} = {a\over b}.\label{eqnk0q0}
\end{equation}

Substituting the expression for $\omega_{k}(q)$ given
by Eq.\ (\ref{eqnwkq}) into Eq.\ (\ref{eqneps}) yields
\begin{eqnarray}
\epsilon(q,\omega_{k}(q)) &=& -a q_1 +
{b\over m}
\Bigl[q_0 k_1 +(k_0-q_0)q_1\Bigr] + O(q_1^2,\omega_1^2,\omega_1 q_1)
\nonumber\\
&=& {b\over m} q_0 k_1 + A k_1 q_1 - B q_1^2 +
{\mbox{\rm cubic\ order\ terms}},
\label{eqnepswkq}
\end{eqnarray}
where $A$ and $B$ are coefficients determined by the
form of the higher-order expansion of $\epsilon(q,\omega_k(q))$.
Note that, by Eq.\ (\ref{eqnk0q0}), the leading order terms in $q_1$
in Eq.\ (\ref{eqnepswkq}) have cancelled, giving the following
leading order behavior in $k_1$ and $q_1$:
\begin{eqnarray}
{\partial \epsilon(q,\omega_k(q))\over\partial q}
&\approx& A k_1 - 2B q_1\nonumber\\
& & \label{leadorder}\\
q_1^* &\propto& \pm k_1^{1/2}.\nonumber
\end{eqnarray}
{}From Eq.\ (\ref{gammapk}) and Eq.\ (\ref{leadorder}),
the asymptotic form of the Born approximation plasmon scattering rate,
as $k\rightarrow k_c$, is
\begin{equation}
\Gamma_{\rm p}(k) \propto  {V_c(q_c)\over q_1^*}
\propto k_1^{-1/2} = (k-k_c)^{-1/2}.
\end{equation}

Note that if up to the $(n-1)$th order term in $q_1$ also cancels away
in Eq.\ (\ref{eqnepswkq}) (which would occur if all derivatives up to
$(n-1)$th order of $\omega_{k_c}(q)$
and $\omega_{\rm p}(q)$ at $q=q_c$ are equal),
then $\epsilon(q,\omega) \sim A k + B q^n$
which leads to a divergence of the plasmon scattering rate which
goes as $\propto k_1^{(n-1)/n}$.
For example, if the curvatures of the curves
$\omega_{\rm p}(q)$ and $\omega_{k_c}(q)$
at $q=q_c$ were the same, then one would have a stronger divergence
$(k-k_c)^{-2/3}$ in $\Gamma_{{\rm p}(k)}$.

This one-sided divergence in the scattering rate indicates that
as the bias voltage between the emitter and the base, $V_{\rm eb}$,
is increased so that the $k$ of the injected electrons increases
above $k_c$ (or alternatively,
if $n$ is decreased so that $k_c$ falls below $k$),
the jump in the scattering rate should be spectacular.
Thus, when $V_{\rm eb}$ is increased through this scattering threshold, the
current passing from emitter to collector should fall dramatically.
Thermal and impurity effects will broaden the divergence in the
scattering rate, but, as can be seen in Fig.\ \ref{fig14},
this effect persists up to relatively high temperatures.

A divergence in the scattering rate at
the optic-phonon emission threshold
has also been predicted\cite{shockley}
in undoped one-dimension systems,
due to the bandedge $E^{-1/2}$ divergence of the density of final states
to which the electrons are scattered.
In the case we have discussed,
the density of final states is finite (because the electrons are not
scattered to the bottom of the band), but there is an inverse square-root
(or possibly stronger) divergence in the
{\em joint} density of states caused by the coupling of the
initial and final states via plasmon emission.
The advantage of a device based on plasmon emission over
one based on optic-phonon emission is that the
threshold energy in the former can be tuned by changing the
density of the doping in the base region, whereas in the latter it
is fixed.  In GaAs, optic-phonon emission threshold energy is $36\,{\rm meV}$,
but at present, the narrowest quantum wires have
an energy separation between the first and second subbands on the order of
about $5\,{\rm meV}$.\cite{goni}  It will take considerable technical
innovation before one has good quality quantum wires in which the
subband separation is larger than the optic phonon energy.
One more problem with utilizing the
optic-phonon principle is that the final states at the bottom of the
band must be unoccupied,
and therefore one cannot dope the system heavily, making it difficult
to make contacts and use it as a three-terminal device.

As the temperature is increased, the divergence becomes a finite peak.
This is because the plasmon line broadens due to Landau damping
(the collective excitations of the system can have a finite life-time
because they can lose energy to the single particles of the system).
The shift of the peak is due to an upward shift in energy of
the plasmon dispersion curve with increasing temperature,
which is a well-known phenomenon in plasma physics.\cite{plasma}
In one-dimension, using the Vlasov formulation\cite{plasma}
(which reduces to the classical limit of the RPA
for equilibrium plasmas), the explicitly upward shift in the plasmon
dispersion with increasing temperature is given, for small $q$, by
\begin{eqnarray}
\omega^2(q) &\approx& {n q^2 V_c(q)\over m}
(1 + {3 m<v^2> \over n V_c(q)}),\nonumber
\\ & & \\
<v^2> &=& {v_F^2\over 3} \Bigl[ 1 + {\pi^2 \over 4}\Bigl({k_{\rm B}T\over
E_{\rm F}}\Bigr)^2 \Bigr] + O(T^4),\nonumber
\end{eqnarray}
where $<v^2>$ denotes the average $v^2$
over the distribution of the electron gas in the base.
The sharp drop in the mean free path
persists to relatively high temperatures (here on the order of tens of
degrees for the parameters chosen), and therefore should be
experimentally observable.  We believe that this sharp drop in the
inelastic mean free path should produce a large NDR in quantum wires as
the injected electrons pass through the threshold energy.

We conclude by mentioning that in higher dimensional
electron systems there is a plasmon threshold as well at the onset of
plasmon emission.  The effect in higher dimensions,
however, is not dramatic because
the ideal mean free path does not change from being infinite below the
threshold to zero above (as it does in the one-dimensional system) since
single particle scattering contributes in higher dimensions, in contrast
to one dimension.  Thus, our proposed NDR in quantum wires is a specific
one-dimensional many-body property.  We also mention that
a doping density of approximately $10^{6}\,{\rm cm}^{-1}$ implies
that a base region of, say, $3000\,{\rm \AA}$
has only $30$ electrons, which may cast doubt on the validity of our
theory.  However, various simulations of
very small number of particles (on the order of 10) in interacting
systems (for example, in the fractional quantum Hall regime or the
Hubbard model) are sufficient to show collective effects,
and therefore it is entirely plausible that the predicted effects
will be seen even with a relatively small number of electrons in the base.
Furthermore, with further improvements in technology, the width of
the wires will decrease, leading to an increase in the allowable doping
density and the number of particles in the base region.

\section{PLASMONS IN QUANTUM WIRES}
\label{sec:sec5}

It is apparent from the foregoing discussions that the
behavior of the plasmons, especially in the presence of
scattering and at finite temperatures, play an
important r\^ole in the many-body physics of
electrons in quantum wires.
Therefore, it seems appropriate for us to conclude this paper with a
brief discussion within the random phase approximation
of the effects of impurities and finite temperature
on the plasmon dispersion in quantum wires.

The RPA plasmon dispersion is given by\cite{mahan,pines} the vanishing of
the dielectric function $\epsilon(q,\omega) = 1 - V_c(q)\Pi_0(q,\omega)$.
The full dispersion for a pure one-dimensional system at $T=0$ within the
RPA is\cite{li,qpli}
\begin{equation}
\omega_{\rm p}(q)= A(q) \Bigl[{\omega^2_+(q) - \omega^2_-(q)\over A(q) -
1}\Bigr]^{1/2},
\end{equation}
where $A(q) = \exp[q\pi/mV_c(q)]$.
An expansion\cite{qpli} to second order in $q/k_F$ gives,
\begin{equation}
\omega_{\rm p}(q) =
|q| \Bigl[v_F^2 + {2\over\pi} v_F V_c(q)\Bigr]^{1/2} +
O(q^3),
\end{equation}
which agrees exactly with the dispersion of the elementary
excitations in the Tomonaga-Luttinger model, a point which we return to later.
For any reasonable confinement model $V_c(q\rightarrow 0)\sim |\ln(qa)|$
and, therefore, $\omega_{\rm p}(q) \sim |q|\;|\ln(qa)|^{1/2}$
in one dimension.
The fact that these one-dimensional plasmons are ungapped at
$\omega\rightarrow 0$ is simply a consequence that the Coulomb
interaction does not provide long-ranged restoring forces for
charge density perturbations in one-dimension.

In this section, we calculate the plasmon dispersion for a
one-dimensional quantum wire for an impure system and at finite
temperatures, by numerically finding the zeros of the dielectric
function on the complex frequency plane, {\em i.e.},
the complex frequencies $\omega_{\rm p}(q)$ such that
$\epsilon(q,\omega_{\rm p}(q))=0$.  The ${\rm Re}[\omega_{\rm p}(q)]$
gives the frequency of plasma oscillation, while the ${\rm Im}[\omega_{\rm
p}(q)]$ gives the damping rate of the mode.
As in the previous sections,
we use the $V(q)$ for electrons confined in an infinite
square well of width $a$, and the RPA form of the dielectric function,
modified by the Mermin formula to include effects of impurities.
Since the modes oscillate as $e^{-i\omega_{\rm p}(q)t}$, the condition of
stability of the system implies that ${\rm Im}[\omega_{\rm p}(q)]\le 0$,
and thus we search for the zeros of $\epsilon(q,z)$ on the
lower-half complex frequency plane.
Note that for ${\rm Im}[z] < 0$, $\epsilon(q,z)$
is given by the {\em analytic continuation} from the upper-half
to the lower-half complex frequency plane.\cite{landau}
Thus, the polarizability in the RPA goes as
\begin{eqnarray}
\Pi_0(q,z) &=& {m\over(\hbar^2)q\pi}
\biggl[\int_{-\infty}^\infty
dk\ {f_{\rm eq}(k+q/2)- f_{\rm eq}(k-q/2)\over k-z/q}\nonumber\\
&&\qquad + 2\pi i \Bigl(f_{\rm eq}(z/q+q/2)-f_{\rm eq}(z/q-q/2)\Bigr)
\theta(-{\rm Im}[z])\biggr],\label{pi0qz}
\end{eqnarray}
where the second term in Eq.\ (\ref{pi0qz}) is required for the
analytic continuation for $\Pi_0(q,z)$ on the lower-half complex plane.

As mentioned in section \ref{sec:sec2}, the presence of impurity scattering
within the wire causes electrons to be diffusive at long times and
large distances.  This behavior of the electrons
overdamps the plasmons and causes them to disappear at small $q$,
as in the two dimensional case,\cite{giuliani}
and this disappearance of the plasmon spectral weight
at small $q$ has important consequences for the
the characteristics of the Fermi surface of a one-dimensional electron gas,
as we have shown in section \ref{sec:sec2}.
Here, we explicitly show in Fig.\ \ref{fig15}(a)
this disappearance by calculating the
plasmon dispersion for a one-dimensional quantum wire in the presence
of impurity scattering.  In the inset of Fig.\ \ref{fig15}(a), we show the
experimental one-dimensional plasmon dispersion compared with the RPA
theory.
We trapped the zeros of the dielectric
function by using the simplex method\cite{vetterling} to search for
the minima of $|\epsilon(q,z)|^2$.
The real and imaginary parts of $\omega_{\rm p}(q)$
are plotted above and below the $x$-axis, respectively.
One can clearly see the overdamping of the plasmon mode
[{\em i.e.}, ${\rm Re}[\omega_{\rm p}(q)] = 0$] as $q$ goes to zero
at a critical wavevector $q_{\rm d}$.
With increasing impurity scattering, $q_{\rm d}$ increases,
as can be seen from the expansion of the dielectric function.
For small $q$ and $\omega$, the polarizability within the
Mermin formula can be expanded to yield
\begin{equation}
\Pi(q,\omega) \approx {nq^2\over m\omega(\omega+i\gamma)}.
\label{hifreqpi}
\end{equation}
This form can also be obtained by the standard replacement of
$\omega^2$ by $\omega(\omega + i\gamma)$
in the denominator of the usual high-frequency result
for the pure system.\cite{pines}
With this form of Eq.\ (\ref{hifreqpi}), the plasmon dispersion is
given by
\begin{equation}
\omega(q) \sim -i{\gamma\over 2} \pm \sqrt{-{\gamma^2\over 4} +
{nq^2\over m} V(q)}.
\end{equation}
{}From Eq.\ (\ref{hifreqpi}), the critical $q_{\rm d}$ at which
$\omega$ becomes completely imaginary is
\begin{equation}
q_{\rm d} = {K\gamma \over  |\ln(K a \gamma)|},
\end{equation}
where $K = \sqrt{m\epsilon_0/(8ne^2)}$.
In the overdamped region, ${\rm Im}[\omega_{\rm p}(q)]$
{\em decreases} with increasing $\gamma$ because a larger $\gamma$
suppresses the electron diffusion rate, slowing the relaxation of the
long-wavelength density perturbations.

We next examine the effect of finite temperature on the
plasmon dispersion.  Finite temperature effects are included by
calculating $\Pi_0(q,\omega)$ at $T\ne 0$, using Eq.\ (\ref{pi0qz}) above,
where the $f_{\rm eq}(k)$ is given by the Fermi functions at temperature $T$.
The results of our calculations are shown in Fig.\ \ref{fig15}(b).
In contrast to the case where impurity scattering exists,
at finite temperatures, the plasmon mode at low $q$ remains undamped.
This is because Landau damping is caused by the transfer of energy
from the collective mode to particles traveling at the same velocity
as the (phase) velocity of the the collective mode, and, therefore,
significant Landau damping at a wavevector $q$ can occur
only if $f_{\rm eq}(\omega_{\rm p}(q)/q)$ is non-negligible; {\em i.e.},
$\omega_{\rm p}(q)/q$ must be less than or on the order of the thermal
or Fermi velocity (whichever is larger).
At small $q$, the plasmon phase velocity
$\omega_{\rm p}(q)/q\sim |\ln(qa)|^{1/2}\rightarrow\infty$, and hence Landau
damping is negligible in this limit.  As $q$ increases, the phase velocity
of the plasmon decreases and hence Landau damping increases.

Before we conclude this section, we argue that the plasmons
that we are have described above (which are associated with the
vanishing of the dielectric function) are {\em precisely} equivalent to the
Tomonaga-Luttinger bosons which are the exact elementary excitations of
the Tomonaga-Luttinger model.   In fact,
as mentioned earlier, the dispersion of the Tomonaga-Luttinger
bosons to order $q^3$ is
{\em exactly} the same as the plasmons calculated within the
RPA,\cite{qpli}
a fact which may be attributed to the fact that only RPA-type
Feynman diagrams ({\em i.e.} the bubble or ring diagrams)
do not cancel out in the Tomonaga-Luttinger model.
As further evidence of the equivalence of the Tomonaga-Luttinger
bosons and what we call plasmons, we note the similarity of the underlying
physical notions of both concepts.
Plasmons are simply long-lived
charge density excitations of the one-dimensional electron gas.
Since Tomonaga-Luttinger bosons are created theoretically by acting on
the ground-state wavefunction by charge density operators,
they are also elementary excitations caused by charge density perturbations.
Therefore, both ``Tomonaga-Luttinger bosons" and ``plasmons" are
simply different names ascribed to the same physical phenomenon.
This may explain why the experimental
results\cite{goni} for one-dimensional plasmon dispersion
(Fig.\ \ref{fig15}(a) inset) agree so well
with the RPA theory, which generally is regarded to be a poor
approximation in lower dimensions.

A similar correspondence between the spin excitations of the
Tomonaga-Luttinger model and that of the single-particle spin
density continuum can also be made.  In the Tomonaga-Luttinger model,
with no spin-dependent interactions, the spin-density excitations have
a linear dispersion $\omega = v_F q$.
The spin-density excitations
(described by the imaginary part of the spin--spin correlation function,
which, within the RPA, is exactly equivalent to the irreducible charge-density
polarizability $\Pi_0(q,\omega)$) also have a dispersion of
$\omega = v_F q$ in the limit that $q\rightarrow 0$.
Unlike the single-particle charge excitations, which have a vanishing
spectral weight as $q\rightarrow 0$ for the RPA in one-dimension (all
the spectral weight is pushed up to the plasmon [Tomonaga boson]
excitations), the spin-density excitations have a nonvanishing spectral
weight in the small $q$ limit.
This is because the spectral weight of the charge excitations
is given by the screened (reducible) polarizability,
$\Pi_0(q,\omega)/\epsilon(q,\omega)$, which is screened out
at small $q$ by the dielectric function,
whereas the spectral weight of the spin excitations is given by the
unscreened (irreducible) polarizability $\Pi_0(q,\omega)$,
which is not suppressed by the screening at small $q$.

\section{SUMMARY}
\label{sec:sec6}

In this paper, we have studied the many-body properties
of semiconductor quantum wires by utilizing a model that is
more realistic than the exactly solvable Tomonaga-Luttinger model,
but which we have solved approximately using the $GW$ approximation,
a renormalized perturbative scheme that has been employed with
success in higher dimensions.
Our approximate calculation provides us an opportunity to
isolate the physical processes which are responsible for the unique
properties of quasi-one-dimensional interacting systems.
In particular, we find that the well-known result that inter-particle
interactions cause the Fermi surface to disappear in one dimension is
due to the virtual excitation of the plasmons at the Fermi surface.
When impurity scattering is included, the Fermi surface
{\em reappears} because the low energy virtual plasmon excitations which
are responsible for the disappearance of the Fermi surface are suppressed
by impurity scattering.  This is consistent with experimental
findings which seem to indicate
the existence of a sharp Fermi surface in semiconductor quantum wires
through large Fermi-edge singularities.\cite{pinczuk}
Furthermore, theoretical studies\cite{lee}
indicate that for a pure Luttinger liquid which is incompressible ({\em i.e.},
where the velocity of the plasmons goes to infinity as $q$ goes to
zero, as is the case studied here) there is no enhancement in the
absorption cross section at the Fermi surface.\cite{enhancement}

While both our model (for zero impurity scattering) and the
Tomonaga-Luttinger model exhibit non-Fermi liquid behavior with the
interacting system having no real Fermi surface, there are important
differences between the two models.  In our model, ${\rm
Im}[\Sigma(k_F,\omega)] \sim |\omega|\,|\ln|\omega||^{1/2}$ for small
$\omega$ whereas in the Tomonaga-Luttinger model, ${\rm
Im}[\Sigma(k_F,\omega) \sim |\omega|^\alpha$, where the parameter
$\alpha$ ($< 1$) depends on the short-range interaction constant of the
model.  The momentum distribution function $n_k$ in our model has a
logarithmic inflection point at $k_F$ whereas in the Tomonaga-Luttinger
model the inflection point in $n_k$ is a power law.
We emphasize that our model is an approximate (leading-order
perturbative) solution of a realistic model of one-dimensional electrons
interacting via the long-range Coulomb interaction whereas the
Tomonaga-Luttinger model is an exact solution of an artificial model of
electrons with linear energy dispersion and infinite band width
interacting via a short-range interaction (in particular, an interaction
which is finite at $q=0$).  An important result of our calculation is
that in the presence of impurity scattering the two models produce
essentially numerically indistinguishable results.

We have also calculated the finite-temperature self-energy
of a semiconductor quantum wire.
To perform this calculation in a numerically efficient manner,
we have developed a formalism for directly analytically continuing
the thermal Matsubara self-energy to real frequencies.
Applying this method to the $GW$ approximation,
we obtain the finite-temperature generalization of the ``line and pole"
decomposition, which in which the real-frequency integration in the
spectral representation is converted into a more computationally efficient
imaginary-frequency sum.

In the absence of impurity scattering and at $T=0$,
the leading order inelastic electron--electron scattering rate
is zero for electrons with energies below the threshold energy for
emission of plasmons but diverges at the plasmon threshold itself.
The divergence as one approaches the threshold from above is
generally inverse square-root, and is due to the fact that the joint density
of states of the electron states connected by the plasmon emission
has the inverse square-root divergence (there are very special circumstances
when the joint density of states is larger, and the divergence is
in fact stronger).  This sudden onset of a very large
$\Gamma_{\rm t}(k)$, which survives both the effects of impurity
scattering and finite temperature, can, in principle, be used in a
a one-dimensional THETA-type device with the possibility of a
large negative differential resistance.
This characteristic has potential applications in switching devices
or oscillators.

We have calculated the plasmon dispersion, including the imaginary
part which describes the damping of the plasmons, for impure wires
and at finite temperatures.  Impurity scattering causes the
plasmons to overdamp and disappear at small $q$,
but finite temperature does not because the phase velocity of
the plasmons diverges as $q\rightarrow 0$.

We also calculated the bandgap renormalization at zero and finite
temperatures. It is found to be on the order of $10$ --
$20\,{\rm meV}$ for typical experimental parameters, which
are consistent with the currently available\cite{cingolani}
experimental results (which have, albeit, multiple-subband occupancy).
We find that the band-gap renormalization can change significantly
between $T = 0\;{\rm K}$ and $300\;{\rm K}$ for wires with very
low electron densities.

\acknowledgments

We thank N. E. Bonesteel, T. Giamarchi, Q. P. Li, Ross ``Bruce"
McKenzie, S. K. Sarker and C. A. Stafford for useful conversations.
This work is supported by the ARO, the ONR and the NSF.

\appendix{Formalism for self-energy
$\Sigma_{\lowercase{ij}} ({\lowercase{k}},{\lowercase{i}}
\nu_{\lowercase{n}})$\\
for the case of many subbands}
\label{app:appa}

In this appendix, we give the formalism for calculating the
self-energy $\Sigma_{ij}(k,i\nu_n)$ for the case where several subbands
are relevant.\cite{vinter}
In this case, the interaction matrix elements and the Green's functions
are labeled by energy subband indices corresponding to the
incoming and outgoing electron lines,
in addition to having the momentum label for the $x$-direction.
(We employ the thermal Green function formalism in this appendix.)

When several subbands are relevant, the Coulomb matrix element
has indices relating to the transitions from different subbands.
We define $V_{ij\,lm}(q)$ as the scattering matrix element for
$i\rightarrow j$ and $l\rightarrow m$, given by
\begin{equation}
V_{ij\,lm}(q)=  \int_{-\infty}^\infty\! dy\!
\int_{-\infty}^\infty\! dy'\; v(q,y-y')\phi_i(y)\phi_j^*(y)
\phi_l(y')\phi_m^*(y'),\label{vijlm}
\end{equation}
where $v(q,y-y')$ is given by Eq.\ (\ref{vqyy}).

The Dyson equation is written as a matrix equation
\begin{equation}
\underline{G}(k,i\nu_n) = \underline{G}^0(k,i\nu_n) +
\underline{G}^0(k,i\nu_n)\,
\underline{\Sigma}(k,i\nu_n)\,\underline{G}(k,i\nu_n)
\end{equation}
where the underline indicates the fact the the Green functions and
self-energies are matrices and the products are matrix multiplications.
The $G_{ij}^0$ is the bare noninteracting Green function
\begin{equation}
G_{ij}^0(k,i\nu_n) = {\delta_{ij}\over i\nu_n - \xi_k - E_j},
\end{equation}
where $E_j$ is the subband energy.
Solving Dyson's equation gives
\begin{equation}
\underline{G} = \Bigl[(\underline{G}^0)^{-1} - \underline{\Sigma}
\Bigr]^{-1}.
\end{equation}
The quasiparticle energies are then given by the poles of
$\underline{G}$; {\em i.e.}, at the frequencies such that ${\rm det}
[(\underline{G}^0)^{-1} - \underline{\Sigma}] = 0$.

The self-energy in the $GW$ approximation\cite{hedin} which is
attached to external electron lines from subbands $i$ and $j$ is
given by
\begin{equation}
\Sigma_{ij}(k,i\nu_n) = -T\sum_l\sum_m
\int {dq\over 2\pi} W_{il\,lj}(q,i\nu_m)\;
G^0_{ll}(k+q,i\nu_n + i\nu_m)
\label{sigij}\end{equation}
where $W_{ij\,lm} = <il|\hat W|jm>$
is the screened interaction between scattering states
$i\rightarrow j$, and  $l\rightarrow m$, and the frequency summation is
over the Bose frequencies $i\nu_m = i2\pi m T$.

To calculate the components of the self-energy one must know the
components of the screened Coulomb interaction $\underline{W}$.
Dyson's equation for the screened interaction,
for the random phase approximation (shown in Fig.\ \ref{fig1}(b)),
which translates to
\begin{equation}
\underline{W} = \underline{V} + \underline{V}\,\underline{\cal
P}\,\underline{W},
\label{u2w}
\end{equation}
where $\underline{\cal P}$ is the bare polarizability (given by the bubble
in the random phase approximation).
[Note that in Eq.\ (\ref{u2w}),
the matrix multiplication implies summation over two
of the four indices, {\em i.e.}, in $W_{ij\, lm}$ the ``row" index is $i$ {\sl
and} $j$, and the ``column" index is $m$ {\em and} $n$, and therefore
the multiplication implied by, say, $\underline{X} =
\underline{W}\;\underline{V}$, is
$X_{ij\, mn} = \sum_{\mu\nu} W_{ij\, \mu\nu} V_{\mu\nu\, mn}$.]
We can solve for $\underline{W}$ from Eq.\ (\ref{u2w}), yielding
\begin{equation}
\underline{W} = \underline{\epsilon}^{-1}\; \underline{V},
\end{equation}
where
\begin{equation}
\underline{\epsilon} = \underline{1} - \underline V\;\underline{\cal P}
\end{equation}
is the dielectric matrix.  Hence, to find $\underline{W}$,
we must invert the dielectric matrix.

The components of the bare polarizability matrix are given by
\begin{equation}
{\cal P}_{ij\,lm}(q,i\nu_n) = \delta_{il}\delta_{jm} \Pi_{lm}(q,i\nu_n)
\end{equation}
where, in the random phase approximation,
\begin{equation}
\Pi_{lm}(q,i\nu_n) =   2\int {dk\over 2\pi}
{f_l(k+q) - f_m(k)\over \xi_{k+q} + E_l - \xi_{k} - E_m - i\nu_n}
\end{equation}
where $f_m(k)$ is the equilibrium Fermi distribution in subband $m$.
Therefore, the components of the dielectric matrix are
\begin{equation}
\epsilon_{ij\,lm}(q,i\nu_n) = \delta_{il} \delta_{jm} -
V_{ij\,lm}(q) \Pi_{lm}(q,i\nu_n).
\end{equation}
Thus, all the ingredients for obtaining $\underline{W}$
are available, and computing its components is simply a matter of algebra.

We now give of the expressions for the $ W_{ij\,lm}(q,i\nu_n)$
for cases where (i) there is one relevant subband, and,
(ii) there are two relevant subbands in a symmetric wire.
In a symmetric wire, the transverse wavefunctions
$\phi_n(y)$ are symmetric (antisymmetric) about $y=0$ for even (odd) $n$,
and since the Coulomb potential $v(q,y)$ is symmetric about $y$,
Eq.\ (\ref{vijlm}) leads to the following symmetry relationships:
\begin{equation}
V_{ij\,lm} = 0\ \ \ {\rm if}\ i+j+l+m = {\rm odd}
\label{symrel1}\end{equation}
and
\begin{equation}
V_{ij\,lm} = V_{ji\,lm} = V_{ij\,ml} = V_{ji\,ml}
= V_{lm\,ij} = V_{ml\,ij} = V_{lm\,ji} = V_{ml\,ji}
\label{symrel2}\end{equation}
({\em i.e.}, the ordering of the indices in the first two or the last
two, or the ordering of the pair of indices, does not matter).
These symmetry relations simplify computations significantly.

In the first case, when the only one relevant subband, as in the case
discussed in the main text, then all equations are scalar equation,
and the screened interaction
is (as in the translationally invariant 3-dimensional case)
\begin{equation}
W_{11\,11}(q,i\nu_n)=
\epsilon^{-1}_{11\,11}(q,i\nu_n)\;V_{11\,11}(q) =
{V_{11\,11}(q)\over 1 - V_{11\,11}(q)\Pi_{11}(q,i\nu_n)}.
\end{equation}

In the second case, where there are two relevant subbands for a
wire that symmetric about the $y$-axis, the dielectric matrix is
\begin{equation}
\underline{\epsilon} = \bordermatrix{&11&22&12&21\cr
		11&1-V_{11\,11}\Pi_{11}& -V_{11\,22}\Pi_{22}&0&0\cr
		22& -V_{11\,22}\Pi_{11}&1-V_{22\,22}\Pi_{22}&0&0\cr
		12&0&0&1-V_{12\,12}\Pi_{12}& -V_{12\,12}\Pi_{21}\cr
		21&0&0& -V_{12\,12}\Pi_{12}&1-V_{12\,12}\Pi_{21}\cr},
\end{equation}
where we have used the symmetry relations of Eqs.\ (\ref{symrel1})
and (\ref{symrel2}).

Taking the inverse of $\underline\epsilon$ gives
\begin{equation}
\underline{\epsilon}^{-1} = \bordermatrix{&11&22&12&21\cr
       11&{1-V_{22\,22}\Pi_{22}\over\epsilon_{\rm intra}}
	 &{V_{11\,22}\Pi_{22}\over\epsilon_{\rm intra}}&0&0\cr
       22&{V_{11\,22}\Pi_{11}\over\epsilon_{\rm intra}}
         &{1-V_{11\,11}\Pi_{11}\over\epsilon_{\rm intra}}&0&0\cr
       12&0&0&{1-V_{12\,12}\Pi_{21}\over\epsilon_{\rm inter}}&
         {V_{12\,12}\Pi_{21}\over\epsilon_{\rm inter}}\cr
       21&0&0&{V_{12\,12}\Pi_{12}\over\epsilon_{\rm inter}}&
	 {1-V_{12\,12}\Pi_{12}\over\epsilon_{\rm inter}}\cr}
\end{equation}
where
\begin{eqnarray}
\epsilon_{\rm intra}(q,i\nu_n) &=& \Bigl(1 - V_{11\,11}(q)
\Pi_{11}(q,i\nu_n)\Bigr)\Bigl(1-V_{22\,22}(q)\Pi_{22}(q,i\nu_n)\Bigr)
\nonumber\\
&& \qquad\qquad\qquad - V_{11\,22}^2(q)
\Pi_{11}(q,i\nu_n)\Pi_{22}(q,i\nu_n)\\
\epsilon_{\rm inter}(q,i\nu_n) &=& 1 - V_{12\,12}(q)\Bigl(\Pi_{12}(q,i\nu_n)
+ \Pi_{21}(q,i\nu_n)\Bigr)\nonumber
\end{eqnarray}
correspond to the intra-level and inter-level dielectric functions,
respectively (the zeros of which give the collective modes of
the systems.\cite{li})
The screened interaction matrix is given by
\begin{equation}
\underline{W} = \underline{\epsilon}^{-1}\underline{V} =
\bordermatrix{&11&22&12&21\cr
	11&{V_{11\,11}(1-V_{22\,22}\Pi_{22})+ V_{11\,22}\Pi_{22}
	    \over \epsilon_{\rm intra}}
	  &{V_{11\,22}\over\epsilon_{\rm intra}}&0&0\cr
	22&{V_{11\,22}\over\epsilon_{\rm intra}}&
	   {V_{22\,22}(1-V_{11\,11}\Pi_{11})+\Pi_{11}V_{11\,22}
	   \over\epsilon_{\rm intra}}
	   &0&0\cr
	12&0&0&{V_{12\,12}\over\epsilon_{\rm inter}}&
	       {V_{12\,12}\over\epsilon_{\rm inter}}\cr
        21&0&0&{V_{12\,12}\over\epsilon_{\rm inter}}&
	       {V_{12\,12}\over\epsilon_{\rm inter}}\cr},
\end{equation}
Eq.\ (\ref{sigij}) is then used to obtain the self-energy
$\Sigma_{ij}(k,i\nu_n)$.

We now briefly discuss the effect of the second subband in the case
where only the first subband is occupied.  Then $\Pi_{22}= 0$, and
\begin{eqnarray}
\Sigma_{11}(k,i\nu_n) &=& -T \sum_{m} \int {dq\over 2\pi}
\biggl({V_{11\,11}(q) G_{11}^0(k+q,i\nu_n + i\nu_m)\over
1 - V_{11\,11}(q)\Pi_{11}(q,i\nu_m)}\nonumber\\
&+& {V_{12\,12}(q) G_{22}^0(k+q,i\nu_n + i\nu_m)\over
1 - V_{12\,12}(q)(\Pi_{12}(q,i\nu_m) + \Pi_{21}(q,i\nu_m))}\biggr).
\label{sig11}
\end{eqnarray}
The first term in the integrand of Eq.\ (\ref{sig11}) is exactly the same
as the expression we used in the main text to calculate the self-energy,
when we ignored the higher subband contribution.  The second term is
the additional contribution of the second subband.  The denominator
of the second term is the intersubband dielectric function,\cite{li}
which only becomes significant for energies on the order of the
intersubband energy separation.  Thus, for the processes in which
the energy is much less than the subband energy separation, as we
have assumed in this paper, the higher subband contribution is
negligible and can be ignored.

\appendix{Calculation for $\Sigma({\lowercase{k}}_F,\omega=
\xi_{\lowercase{k}_F})$ \\ for second order diagram}
\label{app:appb}

We show that for a short-ranged interaction, the second order
diagram shown in Fig.\ \ref{fig1}(c) evaluated at $k_F$ and $\omega=0$
is identically zero.

For $V_c(q) = V_0$ expression for the self-energy is\cite{mahan}
\begin{eqnarray}
\Sigma(k,\omega) &=&
-{V_0^2\over(2\pi)^2}\int dq\int dq'  {1\over q q'}
\Bigl(n_F(\xi_{k+q'})[n_F(\xi_{k+q}) - n_F(\xi_{k+q+q'})] \nonumber\\
&&+ n_F(\xi_{k+q+q'})(1-n_F(\xi_{k+q}))\Bigr)
\end{eqnarray}
The Fermi functions restrict the four integration regions to the $q,q'$
space shown in Fig.\ \ref{fig16}.
We calculate the contribution for each region.

For regions I, II, and III, we obtain
\begin{eqnarray}
{\rm I} &=&\lim_{\varepsilon\rightarrow 0}\int_0^\varepsilon {dq\over q}
\int_{\varepsilon-q}^{\varepsilon} {dq'\over q'}
=\int_0^1 {dx\over x} \int_{1-x}^1 {dx'\over x'}\nonumber\\
&=&-\int_0^1 dx {\ln(1-x)\over x} = {\pi^2\over 6}\nonumber\\
{\rm II}
&=&\int_0^{2k_F} {dq\over q} \int_{2k_F-q}^{2k_F} {dq'\over q'}\\
&=&\int_0^1 {dx\over x'}\int_{1-x}^1 {dx'\over x'} = {\pi^2\over 6}
\nonumber\\
{\rm III}
&=&\int_{-\infty}^{-2k_F} {dq\over q}\int_{-q-2k_F}^{-q} {dq'\over q'}
=\int_{-\infty}^{-1} {dy\over y}\int_{-y-1}^{-y} {dy'\over y'}\nonumber\\
&=&\int_{-\infty}^{-1} {dy\over y} \ln\Bigl({y\over 1+y}\Bigr)
=\int_{0}^1 dx {\ln(1-x)\over x} = -{\pi^2\over 6}\nonumber\\
\end{eqnarray}
A calculation similar to that of region III also gives a contribution
of $-\pi^2/6$ from region IV.  Therefore, the second order contribution to
$\Sigma(k_F,\omega=0)$, which is given by the sum of the
contributions from I, II, III and IV, is identically zero.

\appendix{Form of $[\Sigma({\lowercase{k}}_F,\omega)]$
as $\omega\rightarrow 0$ in the\\ $GW$ and $RPA$ approximation in
one dimension}
\label{app:appc}

In this appendix, we give the details of the calculation yielding
the $\omega\rightarrow 0$ forms of the imaginary part of the retarded
self-energy in the one-dimensional electron gas, within the $GW$
approximation, both with and without impurities.

First, we define two useful energy scales
\begin{equation}
E_{\rm F} = {k_F^2\over 2m},
\qquad\qquad\qquad\qquad
E_{\rm pot} =  {n e^2\over\epsilon_0} = {2e^2 k_F\over(\pi\epsilon_0)}.
\end{equation}
$E_{\rm F}$ is the Fermi energy, and $E_{\rm pot}$ is the potential
energy for two electrons separated by the average distance between electrons
at density $n$.  As in higher dimensions, we define the ratio of
these energies as\cite{define}
\begin{equation}
r_s = {4 e^2 \over\pi \epsilon_0 v_F}
= {E_{\rm pot}\over E_{\rm F}},
\end{equation}
where $v_F = k_F/m$ is the Fermi velocity.

As mentioned previously, the only contribution to the
imaginary part in the $GW$ approximation comes from $\Sigma_{\rm pole}$.
The contribution to ${\rm Im}[\Sigma]$ comes from two
sources: (i) the single particle continuum
where ${\rm Im}[\Pi(q,\omega)] \ne 0$, and from (ii)
the plasmon contribution ({\em i.e.}, when the ${\rm Re}[\epsilon(q,\omega)]=
0$
and $|{\rm Im}[\epsilon(q,\omega)| = 0^+$).
The $\theta$-functions in Eq.\ (\ref{imsigma})
dictate that the only contribution
to ${\rm Im}[\Sigma]$ occurs when $\omega-\xi_{k+q}$ and
$\xi_{k+q}$ have different signs.
For $k = k_F$, $\xi_{k_F+q} =  q^2/(2m)+qv_F$, and hence
for small $\omega$, where one can linearize the energy-momentum
curves about $q=0$ and $q=-2k_F$, yielding
\begin{eqnarray}
{\rm Im}[\Sigma(k_F,\omega)] &\approx& \int_0^{\omega/v_F} {dq\over 2\pi}
\;\; {\rm Im}\Bigl[{V_c(q)\over \epsilon(q,qv_F-\omega)}\Bigr]\nonumber\\
&&+ \int_{-\omega/v_F}^{0} {dq'\over 2\pi}
\;\;{\rm Im}\Bigl[{V_c(-2k_F+q')\over
\epsilon(-2k_F+q', -q'v_F-\omega)}\Bigr].\label{C1}
\end{eqnarray}
The bold lines in Fig.\ \ref{fig2} show the energy-momentum curves over
which the there is a contribution for the self-energy.
We show below that, as $\omega \rightarrow 0$, the
${\rm Im}[\Sigma]$ is dominated by the {\em plasmon} contribution
in the case of clean wires, and this contribution leads to a
limiting form of $|{\rm Im}[\Sigma(k_F,\omega)]|
\sim |\omega|\, |\log(|\omega|)|^{1/2}|$.

\subsection{Plasmon Contribution to ${\rm Im}\Sigma(k_F,\omega)$
in a clean system}

The plasmon contribution comes in when the real part of $\epsilon$
goes through zero (corresponding to the energy-momentum curve cutting
through the plasmon dispersion curve).  The contribution goes as
\begin{equation}
{\rm Im}[\Sigma(k,\omega)] = -
{1\over 2} V_c(q^*)  \Bigl|{\partial\epsilon(q,\xi_{k+q}-\omega)\over
\partial q}\Bigr|_{q=q^*}^{-1}\label{C1a}
\end{equation}
where $q^*$ is the wavevector at the intersection of the
energy curve $\xi_{k+q}-\omega$ and the plasmon dispersion curve
$\omega_{\rm pl}(q)$.

For small $\omega$, the plasmon intersection $q^*$ occurs at small $q$,
and so one can linearize the energy-momentum curve
\begin{equation}
\xi_{k_F+q}-\omega \sim  {k_F q\over m} - \omega
\label{energymom}
\end{equation}
which implies that, for an intersection at $q^*$, the corresponding
energy at which it occurs is $E^* = k_F q^*/m$.

The dispersion relation for the plasmon, $\omega_{\rm pl}(q)$,
is given by the solution of $\epsilon(q,\omega_{\rm pl}(q)) = 0$.
In the random phase approximation in a quasi-one-dimensional system,
the dispersion relation is given by\cite{qpli}
\begin{equation}
\omega_{\rm pl}(q) = qv_{\rm F} \sqrt{1 + {2 V_c(q)\over \pi v_F}}.
\end{equation}
We now study two different cases, the Coulomb interaction (where
$V_c(q)\rightarrow\infty$ as $q\rightarrow 0$), and the short-ranged
interaction (where $V_c(q)$ is finite when $q\rightarrow 0$).
In both cases, we obtain non-Fermi-liquid behavior from the
$GW$ approximation to the self-energy.

\underline{\em Coulomb interaction} ---
For the case of the Coulomb interaction where
$V_c(q) \approx 2 e^2 |\ln|q a||/\epsilon_0$, the plasmon dispersion
for small $q$ is
\begin{equation}
\omega_{\rm pl}(q) \approx q \sqrt{2 v_F V_c(q)\over\pi}
\approx q v_F \sqrt{r_s |\ln|q a|| }.
\label{omegapl}
\end{equation}
The intersection of the energy-momentum curve $\xi_{k_F+q}-\omega$
and the plasmon curve $\omega_{\rm pl}(q)$, from Eq.\
(\ref{energymom}) and (\ref{omegapl}),
occurs at
\begin{equation}
\omega \approx  q^* v_F \Bigr(\sqrt{r_s |\ln|q^* a||}+1\Bigr).
\label{omegaapprox}
\end{equation}
In the limit of small $|\omega|$ and $|q^*(\omega)|$, the $1$ on the
right-hand side of Eq.\ (\ref{omegaapprox}) can be ignored, yielding
\begin{eqnarray}
q^* &\approx& {\omega\over v_{\rm F}}
{1\over\sqrt{r_s|\ln|q^* a||}}\nonumber\\
&\approx& {\omega\over v_{\rm F}}
{1\over\sqrt{r_s|\ln|a\omega/(\sqrt{r_s}v_{\rm F})||
- \ln(\sqrt{|\ln|q^* a||})}}.\label{qstar}
\end{eqnarray}
Since $|q|^*\ll |\omega|$ in the limit $\omega\rightarrow 0$,
the second term in the denominator of the last expression in
Eq.\ (\ref{qstar}) is negligible when compared to the first term
and hence
\begin{equation}
q^*(\omega) \approx {\omega\over v_{\rm F}} {1\over
\sqrt{r_s|\ln|a\omega/(\sqrt{r_s}v_{\rm F})||}}.
\label{C7}\end{equation}

Now the derivative of the dielectric function for $q\rightarrow 0$
and $\omega/q \gg v_F$ is
\begin{eqnarray}
{\partial \epsilon(q,\xi_{k+q}-\omega) \over \partial q}
&\approx& {\partial V_c(q)\Pi(q,\xi_{k+q}-\omega)\over \partial q}
\nonumber\\
&\approx& {\partial V_c(q) \over\partial q} \Pi(q,{qk_F\over m}-\omega)
   + V_c(q){\partial \Pi(q,{qk_F\over m}-\omega)\over\partial q}
\nonumber\\
&=& {n\over m}{\partial V_c(q) \over \partial q} {q^2\over
(qk_F/m-\omega)^2} + {n\over m} V_c(q) {\partial
(q^2/(qk_F/m-\omega)^2)\over \partial q}\nonumber\\
= r_sv_{\rm F}^2\Bigl[{q\over (qk_F/m - \omega)^2} &+&
\ln(|qa|) \Bigl({2 q\over (qk_F/m - \omega)^2} - {2(q^2 k_F/m)\over
(q k_F/m - \omega)^3}\Bigl)\Bigr].
\label{C8}
\end{eqnarray}
Given that $q \sim \omega/|\ln(|\omega|)|\ll \omega$ as
$\omega\rightarrow 0$, the second term in the last expression of Eq.\
(\ref{C8})
dominates, and thus we have
\begin{equation}
\Bigl[{\partial\epsilon(q,\xi_{k+q}-\omega)\over\partial q}
\Bigr]_{q=q^*}
\approx r_sv_{\rm F}^2 |\ln|q^* a|| {2 q^*\over \omega^2}.
\label{C9}
\end{equation}
Substituting Eqs.\ (\ref{C9}) and (\ref{C7}) into Eq.\ (\ref{C1a}) gives
\begin{equation}
{\rm Im}[\Sigma(k_F,\omega)] =
-{\pi\sqrt{r_s}\over 8}|\omega|
\sqrt{\Bigl|\ln\Bigl|{a\omega\over \sqrt{r_s}v_{\rm F}}\Bigr|\Bigr|}.
\end{equation}
Thus we have shown that in the clean limit, within the $GW$
and RPA approximations the imaginary part of the self-energy goes as
\begin{equation}
|{\rm Im}[\Sigma(k_F,\omega)]| \sim |\omega|\sqrt{|\ln|\omega||},
\end{equation}
which implies there is no Fermi surface,
and we have shown that the physical mechanism
for the disappearance of the Fermi surface is the {\em virtual
emission of low energy plasmons}.

\underline{\em Short-ranged interaction} ---
In the case of a short-ranged interaction ({\em i.e.}, a delta-function
in real space) where $V_c(q)=U_0$ is a
constant, we show below that $|{\rm Im}[\Sigma(k_F,\omega)]| \sim |\omega|$
for small $\omega$.

The plasmon dispersion for short-ranged interactions is
\begin{equation}
\omega_{\rm pl}(q) = q v_{\rm F} \sqrt{1 + {2 U_0 \over \pi v_F}}.
\end{equation}
The dielectric function at small $q$ for $z=\xi_{k+q} - \omega$
is given by
\begin{eqnarray}
\epsilon(q,\xi_{k_F+q}-\omega) &\approx& 1 - {m U_0\over \pi q}
\ln \Bigl|{1 + {v_F q^3 \over m (\omega^2 - 2\omega v_F q)}\over
           1 - {v_F q^3 \over m (\omega^2 - 2\omega v_F q)}}\Bigr|
\nonumber\\
&\approx&  1 - {2 U_0 v_F q^2\over \pi(\omega^2 - 2\omega v_F q)}
\end{eqnarray}
and therefore,
\begin{equation}
{\partial\epsilon(q,\xi_{k_F+q}-\omega)\over\partial q} \approx
-{2 U_0 v_F\over\pi} \Bigl[
{2q\over (\omega^2 - 2\omega v_F q)} -
{2 \omega v_F q^2\over (\omega^2 - 2 \omega v_F q)^2}\Bigr].
\label{C14}
\end{equation}
Substituting Eq.\ (\ref{C14}), into Eq.\ (\ref{C1a}) yields the plasmon
contribution to the self-energy, in the limit of $\omega\rightarrow 0$
\begin{equation}
{\rm Im}[\Sigma_{\rm pl}(k_F,\omega)] =
-{\pi (\sqrt{ 1 + 2 U_0/(\pi v_F)} - 1)^2 \omega \over
4 \sqrt{ 1 + 2 U_0/(\pi v_F)} }.
\end{equation}
Thus in the case where the interaction is short-ranged, one
obtains the imaginary part of the self-energy being proportional
to $\omega$, giving a marginal Fermi liquid.  From exact solutions,
we know that short-ranged interactions give non-Fermi liquids.
We have shown that the $GW$ approximation is sufficient to
describe this non-Fermi-liquid behavior for short-ranged interactions.
(Note, however, that the functional dependence is different in the exact
Luttinger liquid situations, where ${\rm Im}[\Sigma(k_F,\omega)] \sim
|\omega|^\alpha$, with $\alpha < 1$.)

\subsection{Single particle contribution to ${\rm Im}[\Sigma(k_F,\omega)]$
in a clean system}

The single-particle excitation contribution to ${\rm Im}[\Sigma]$
comes from each of the two terms on the right-hand-side of Eq.\ (\ref{C1}).
The first term corresponds
to virtual electron-hole excitations around $q=0$.
Within the region of integration
$0 < q < \omega/v_F$, ${\rm Im}[\epsilon] \ne 0$ only in a region
the size of $\sim\omega^2$.  In this region, both the real and imaginary
parts of $\epsilon(q,qv_F-\omega)$ are of the order of
$V_c(\omega/v_F)/\omega$, and therefore the integrand
$V_c(q)\,{\rm Im}[\epsilon^{-1}(q,qv_F-\omega)]$ is on the order
of $\omega$.  Thus, the contribution of this first term is on the
order of the product of the size of region of integration and the magnitude
of the integrand, $\omega^2 \times \omega = \omega^3$.

The contribution from the second term of Eq.\ (\ref{C1}) corresponds to
virtual electron-hole excitations around $q=2k_F$.
Here, the region where ${\rm Im}[\epsilon(2k_F + q', -q'v_F -\omega)]\ne 0$
is of order $\omega$.  In this region, ${\rm Im}[\epsilon]\sim
{\rm constant}$, and ${\rm Re}[\epsilon] \agt |\ln(\omega)|$,
leading to  the integrand being on the order of $|\ln(\omega)|^2$.
Therefore the magnitude of the contribution from this term is the
product of the size of the region of the integration and the
magnitude of the integrand, which is $\omega/|\ln(\omega)|^2$.  Note that the
$\ln|\omega|$ comes from the form of the RPA $\Pi_0(q,\omega)$ and is {\sl
independent} of the form of $V_c(q)$. This term dominates the
$q=0$ term, and hence the single-particle contribution to the self-energy is
\begin{equation}
|{\rm Im}[\Sigma(k_F,\omega)]|\sim {|\omega|\over(\ln|\omega|)^2}.
\end{equation}

\subsection{${\rm Im}[\Sigma(k_F,\omega)]$ as $\omega\rightarrow 0$
for dirty systems}

In the case where there is impurity scattering, electrons are diffusive
and, as shown in section \ref{sec:sec5}, the plasmons no longer exist at small
wavevectors.  Thus, for small $\omega$, there is no plasmon contribution
to the imaginary part of the self-energy.  In this subsection, we show
that the remaining contribution goes as
${\rm Im}[\Sigma(k_F,\omega)] \sim \omega^2$.

We find it convenient to define the following
the following dimensionless variables:
\begin{eqnarray}
K,Q = {k\over k_F},{q\over k_F},
&\qquad\qquad\qquad\qquad&
\Omega = {\omega\over E_F},\nonumber\\
\tilde\gamma = {\gamma\over E_F},
&\qquad\qquad\qquad\qquad&
\tilde V(Q) = {V_c(q)\over 2e^2/\epsilon_0},\\
\tilde\Pi(Q,\Omega) = {\Pi(q,\omega)
\over (2m/\pi k_F)},
&\qquad\qquad\qquad\qquad&
\tilde\Sigma(K,\Omega) = {\Sigma(k,\omega)\over E_{\rm pot}} =
{\Sigma(k,\omega)\over 2e^2 k_F/(\pi\epsilon_0)}.\nonumber
\end{eqnarray}
With these definitions, $\tilde\Pi(Q\rightarrow 0,\Omega=0) = -1$ and
$\epsilon(Q,\Omega) = 1 - r_s \tilde V_c(Q) \tilde\Pi(Q,\Omega).$
In terms of these dimensionless variables, Eq.\ (\ref{C1}) becomes
\begin{eqnarray}
\tilde\Sigma(K = 1,\Omega) &\approx& \int_0^{\Omega/2} {dQ\over 2}
\;\tilde V_c(Q)\, {\rm Im}[\epsilon^{-1}(Q, 2Q-\Omega)]\nonumber\\
&+& \int_{-\Omega/2}^0 {dQ'\over 2}\; \tilde V_c(-2 + Q')\,
{\rm Im}[\epsilon^{-1}(-2 + Q', -2Q'-\Omega)].
\label{C18}
\end{eqnarray}

The polarizability is given by the Mermin form.
In the region $Q,\;\Omega \ll\tilde\gamma$, the polarizability is of the
form
\begin{equation}
\tilde\Pi_\gamma(Q,\Omega') \approx -{Q^2\over Q^2 - i\Omega'\tilde\gamma/4}
\end{equation}
which gives
\begin{equation}
{\rm Im}[{1\over\epsilon(Q,\Omega')}] =  - {1\over 4}
{r_s \Omega' \tilde\gamma Q^2 \tilde V_c(Q) \over
Q^4 ( 1 + r_s \tilde V_c(Q))^2 + (\Omega' \tilde\gamma/4)^2}.
\end{equation}
Therefore, the virtual transitions around $Q=0$ [the first term
in Eq.\ (\ref{C18})] give a contribution
\begin{eqnarray}
\tilde\Sigma_1(K=1,\Omega) &\approx& {r_s\tilde\gamma\over 8}
\int_0^{\Omega/2} dQ {Q^2 (\Omega- 2Q)\tilde V_c^2(Q)
\over Q^4 (1 + r_s \tilde V_c(Q))^2 + ((\Omega - 2 Q)\tilde\gamma/4)^2}
\nonumber\\
&=&{r_s \Omega^2\over 2\tilde\gamma}
\int_0^1 {\tilde V_c^2\Bigl(\Omega (1-y)/2\Bigr) (1-y)^2 y \over
(1-y)^4( (1 + r_s \tilde V_c\Bigl(\Omega(1-y)/2\Bigr))
\Omega/\tilde\gamma)^2 + y^2}.\label{C21}
\end{eqnarray}
As $\Omega\rightarrow 0$, the integral in the last line of
Eq.\ (\ref{C21}) diverges in the region $y=0$, and therefore only the
behavior of the integrand around that region is relevant.  Therefore,
the asymptotic form of $\tilde\Sigma$
(using the asymptotic form $\tilde V_c(Q) \sim |\ln(Q)|$) is
\begin{eqnarray}
\tilde\Sigma_1(K=1,\Omega) &\approx& {r_s\tilde\gamma\over 8}
\int_0^1 dy \; {\ln(\Omega)^2 y\over (\omega^2 r_s^2 \ln(\omega)^2
/\tilde\gamma^2) + y^2}\nonumber\\
&=& {r_s \Omega^2 \ln(\Omega)^2 \over 2\tilde\gamma}
\ln\Bigl( {r_s^2 \Omega^2
\ln(\Omega)^2 \over \tilde\gamma}\Bigr),
\end{eqnarray}
which to leading order, goes as $\Omega^2|\ln|\Omega||^3$.

Now we study the contribution from the $Q=2$ virtual transitions.
The polarizability in this region is
\begin{eqnarray}
\tilde\Pi_{\gamma}(Q,\Omega) &=&
\Pi_0(Q,\Omega=0) - i C(Q)\Omega + 0(\Omega^2),\nonumber
\\ & & \\
C &=& {(\Pi_0(Q,i\tilde\gamma)-\Pi_0(Q,0))
\Pi_0(Q,0)\over \tilde\gamma\,\Pi_0(Q,i\tilde\gamma)}.\nonumber
\end{eqnarray}
Using this expansion,\cite{expand} we obtain
\begin{equation}
{\rm Im}[\epsilon^{-1}(Q,\Omega)] \approx
{{\rm Im}[\Pi_{\gamma} (Q,\Omega)]\over
(r_s\tilde V_c(Q) {\rm Re}[\Pi_0 (Q,0)])^2},
\end{equation}
which yields a contribution to $\tilde\Sigma(K=1,\Omega)$,
from Eq.\ (\ref{C18}), of
\begin{equation}
\tilde\Sigma_2(K=1,\Omega) \sim {\Omega\over |\ln|\Omega||^2}
\int_{-\Omega/2}^0 dQ \sim {\Omega^2\over |\ln|\Omega||^2}.
\end{equation}

Thus, in this case where electrons diffuse due to impurity scattering,
the virtual excitations close to $Q=0$ dominate, and
\begin{equation}
|{\rm Im}\Sigma(k_F,\omega)| \sim \omega^2 |\ln|\omega||^3.
\end{equation}

\subsection{\ Form of ${\rm R{\lowercase{e}}}[\Sigma({\lowercase{k}}_F,
\omega)]$ for clean systems}

The $\omega\rightarrow 0$ limit of the real part of the self-energy
at $k=k_F$ is
\begin{eqnarray}
{\rm Re}[\tilde\Sigma(k_F,\omega)] &\approx& \int_0^{\Omega/2} {dQ\over 2}
\;\tilde V_c(Q)\, {\rm Re}[\epsilon^{-1}(Q, 2Q-\Omega)-1]\nonumber\\
&+& \int_{-\Omega/2}^0 {dQ'\over 2}\; \tilde V_c(-2 + Q')\,
{\rm Re}[\epsilon^{-1}(-2 + Q', -2Q'-\Omega)-1].
\label{C27}
\end{eqnarray}
The real part of $\epsilon^{-1}(Q,\Omega)$
in the integration region $Q=0,\Omega/2$ is well-behaved
(aside from an inconsequential principal part divergence at the
plasmon frequency).  The $\epsilon^{-1} - 1$ term basically is a constant,
and the magnitude of the integrand is determined by the $\tilde V_c(Q)$ term.
Since the size of region of integration goes as $\Omega$ and the integrand goes
approximately as $\tilde V_c(\Omega)\sim|\ln|\Omega||$, the contribution from
the first integral on the right-hand side of Eq.\ (\ref{C27}) is
on the order of $|\Omega \ln|\Omega||$.
In the second integral, the integrand does not diverge, and hence
its contribution is on the order of $\Omega$ and can be neglected
in comparison to the first. Hence
\begin{equation}
{\rm Re}[\Sigma(k_F,\omega)] \sim |\omega\ln|\omega||.
\end{equation}
The slope of ${\rm Re}[\Sigma(k_F,\omega)]$ at $\omega=0$ is infinite
(in fact, this is necessary for $A(k_F,\omega)$ to be integrable), implying
from Eq.\ (\ref{19}) that $Z_F =0$, which is consistent with the fact that
there is no Fermi surface.

\appendix{${\rm I{\lowercase{m}}}[\Sigma({\lowercase{k}}_F,\omega)]$ as
$\omega\rightarrow 0$, in the\\ $GW$ and $RPA$ approximation
in two dimensions}
\label{app:appd}

In this appendix, we calculate, for the sake of comparison,
the imaginary part of the $\Sigma(k_F,\omega)$,
due to plasmon and single particle emission in two dimensions
for the case of Coulomb interaction, which is\cite{ando}
\begin{equation}
V_c(q) = {2 \pi e^2\over\epsilon_0 q}.
\end{equation}

In the $GW$ approximation for a spherically symmetric parabolic band,
the imaginary part of the self-energy can be written as
(for the case of $\omega > 0$)
\begin{equation}
{\rm Im}[\Sigma(k,\omega)] =
{2 m\over (2\pi)^2}\int_0^\omega d\xi \int_0^\pi d\theta\;
V_c\bigl(q(\theta,\xi)\bigr)\;
{\rm Im}[\epsilon^{-1}(q(\theta,\xi),\xi-\omega)]
\label{D3}
\end{equation}
where $\epsilon(q,\omega) = 1 - V_c(q) \Pi(q,\omega)$ and
\begin{equation}
q^2(\theta,\xi)
= k^2 + k_F^2 + 2 m \xi - 2 k \sqrt{ k_F^2 + 2 m E} \cos \theta.
\end{equation}
The exact form of $\Pi(q,\omega)$ in RPA can be found in references
\cite{stern} and \cite{ando}.
We define the reduced variables
\begin{eqnarray}
K,Q = {k\over k_F},{q\over k_F},
&\qquad\qquad\qquad\qquad&
\Omega = {\omega\over E_F},\nonumber\\
y = {2 m\xi\over k_F^2} = {\xi\over E_F},
&\qquad\qquad\qquad\qquad&
\tilde\gamma = {\gamma\over E_F},\nonumber\\
\tilde V_c(Q) = {V_c(q)\over (2\pi e^2/(q_F\epsilon_0)}
= {1\over Q},
&\qquad\qquad\qquad\qquad&
R_s = {2e^2\over\epsilon_0 v_F},\label{D5}\\
\tilde\Sigma(K,\Omega) =
{\Sigma(k,\omega)\over e^2 k_F/(\pi\epsilon_0)},
&\qquad\qquad\qquad\qquad&
\tilde\Pi(Q,\omega) = {\Pi(q,\omega)\over m/\pi}.\nonumber
\end{eqnarray}
With these definitions, $\tilde\Pi(Q\rightarrow 0,\Omega=0) = -1$ and
$\epsilon(Q,\Omega) = 1 - R_s\tilde V_c(Q)\tilde \Pi(Q,\Omega)$.
For the case $k=k_F$,
${Q^2} = 2 + {y} - 2\sqrt{ 1 + y} \cos\theta$, which yields
\begin{equation}
d\theta = {Q\, dQ \over \sin\theta(Q,y)\;\sqrt{1 + y}},
\label{D6}
\end{equation}
where,
\begin{equation}
\sin\theta(Q,y) = \sqrt{ 4 Q^2 - y^2 + 2 y Q^2 - Q^4\over 4 (1+y)}.
\end{equation}
\label{D7}
Therefore, using Eqs.\ (\ref{D5}), (\ref{D6}) and
(\ref{D7}), Eq.\ (\ref{D3}) can be rewritten as
\begin{equation}
{\rm Im}[\tilde\Sigma(k_F,\omega)]=
\int_0^\Omega dy \int_{Q_-(y)}^{Q_+(y)} dQ\;
{\rm Im}[\epsilon^{-1}(Q,y-\Omega)]
{1\over \sqrt{4Q^2 -y^2 +2 yQ^2 -Q^4}}\label{D8}
\end{equation}
where,
\begin{eqnarray}
Q_{-}(y) &=& \Bigl[2+y-2\sqrt{1+y}\Bigr]^{1\over 2}
                     = {y\over 2} + O(y^2),\nonumber
\\ & & \label{D9}\\
Q_{+}(y) &=&\Bigl[2+y+2\sqrt{1+y}\Bigr]^{1\over 2}
	   = 2+{y\over 2} + O(y^2).\nonumber
\end{eqnarray}
Within the $GW$ approximation for two dimensions, Eq.\ (\ref{D8})
is an exact result.
Now, we calculate the plasmon and single-particle excitation
contributions to $\Sigma(k_F,\omega)$ in the $\omega/E_F \ll 1$ limit.

\subsection{Plasmon contribution in two dimensions}

The plasmon contribution to the imaginary part of $\Sigma(k,\omega)$
occurs when ${\rm Re}[\epsilon(Q,y-\Omega)] = 0$ in Eq.\ (\ref{D8}).
Since we are interested in small frequencies, the dielectric function
in two-dimensions is given in the small frequency limit by\cite{pines}
\begin{equation}
\epsilon(Q,y-\Omega) = 1 - {2 R_s Q \over (y-\Omega)^2}.
\end{equation}
and therefore, the wavevector at which ${\rm Re}[\epsilon] = 0$ is
\begin{equation}
Q_{\rm p}(y) = {(y-\Omega)^2\over 2 R_s}.
\label{D11}
\end{equation}
Hence,
\begin{eqnarray}
{\rm Im}[\epsilon^{-1}(Q,y-\Omega)] &=& {\pi \delta\Bigl(Q - Q_{\rm
p}(y)\Bigr)\over |\partial\epsilon(Q,y-\Omega)/\partial Q|}\nonumber\\
&=& \pi \delta\Bigl(Q-Q_{\rm p}(y)\Bigr) {(y-\Omega)^2\over 2 R_s}
\label{D12}
\end{eqnarray}
There is a non-zero plasmon contribution to the integral in Eq.\
(\ref{D8}) only if the delta function peaks between $Q_-(y)$ and $Q_+(y)$;
that is (since $Q_{\rm p}(y)$ is small) only if $Q_{\rm p}(y) > Q_-(y)$.
This occurs if the condition
\begin{equation}
{(\Omega-y)^2\over R_s} > y,
\end{equation}
is satisfied, or equivalently, if
\begin{equation}
y < y_{\rm max} =
{R_s\over 4} \Bigl(1 + {2\Omega\over R_s} - \sqrt{1 +
{4\Omega\over R_s}}\Bigr) =
{\Omega^2\over R_s} + O(\Omega^3).
\label{D14}
\end{equation}
Thus, from Eqs.\ (\ref{D8}), (\ref{D11}), (\ref{D12}) and (\ref{D14}),
the plasmon contribution, to lowest order in $\Omega^2$, is
\begin{eqnarray}
{\rm Im}[\Sigma_{\rm pl}(k_F,\omega)] &\approx&
{e^2 k_F\over 2\epsilon_0 R_s} \int_0^{\Omega^2/R_s} dy
{(y-\Omega)^2\over \sqrt{(\Omega^2/R_s)^2 - y^2}}\nonumber\\
&\approx& {E_F\over 2} \Omega^2 \int_0^1 {dz
\over\sqrt{1-z^2}}\nonumber\\
&=& {\pi\over 4}{\omega^2\over E_F}.
\end{eqnarray}
Therefore, the plasmon contribution to the imaginary part of the
self-energy goes as as $\omega^2$.

Note that the quantity we calculate here, ${\rm Im}[\Sigma(k_F,\omega)]$,
is different from the inverse inelastic lifetime of the quasiparticle,
$\tau^{-1}_{\rm ee}(k) = 2\,{\rm Im}[\Sigma(k_F,\xi_{k_F})]$.
For $\tau^{-1}_{\rm ee}$ there is a minimum critical $|p-p_F|$
before the plasmon contributes to $\tau^{-1}_{\rm ee}$,\cite{giuliani0}
whereas for ${\rm Im}[\Sigma(k_F,\omega)]$, there is no gap.

\subsection{The single-particle excitation contribution}

We show in this subsection that the single-particle excitation contribution
to the imaginary part of the two-dimensional self-energy goes as
$\omega^2 |\ln(|\omega|)|$, and hence gives the dominant contribution
to the Im$[\Sigma]$ in a two-dimensional electron gas.  This result
has been obtained previously,\cite{highdim,giuliani0} in similar contexts,
although details of the actual calculation were never given.
Below, we give a quick derivation of this result.

For small $\Omega$, the polarizability goes as\cite{stern,ando}
\begin{eqnarray}
{\rm Re}[\tilde\Pi(Q,\Omega)] &=& -1\nonumber\\
& &\\ \label{D16}
{\rm Im}[\tilde\Pi(Q,\Omega)] &\approx&
-{\Omega\over Q\sqrt{4-Q^2}}\nonumber
\end{eqnarray}
and hence, since $|{\rm Im}[\Pi(Q,\Omega)]| \ll
|{\rm Re}[\Pi(Q,\Omega)]|$ in this limit,
\begin{equation}
{\rm Im}[\epsilon^{-1}(Q,\Omega)] \approx
{{\rm Im}[\epsilon(Q,\Omega)]\over({\rm Re}[\epsilon(Q,\Omega)])^2} \approx
{R_s\Omega\over \sqrt{4-Q^2}(Q+R_s)^2}.
\end{equation}
The single-particle contribution to the imaginary part of the self-energy,
from Eq.\ (\ref{D9}), is
\begin{equation}
{\rm Im}[\Sigma_{\rm sp}(k_F,\omega)] \approx
{e^2 k_F\over \pi\epsilon_0}\int_0^\Omega dy\; (\Omega-y)
\int_{\sim\Omega}^2 dQ\; {R_s\over Q(Q+R_s)^2(4-Q^2)}
\label{D18}
\end{equation}
where terms higher order in $y$ and $\Omega$ have been ignored.  The
lower limit for the $Q$-integration is given by a cutoff on the
order of $\Omega$, which is where the expansion given in Eq.\ (\ref{D16})
starts to fail (the remaining part of the integral is from $\sim \Omega$
to $Q_-$ is well behaved and gives negligible contribution).
The leading order contribution to the Eq.\ (\ref{D18}) is
\begin{eqnarray}
{\rm Im}[\Sigma(k_F,\omega)] &\approx&
{e^2 k_F\over\pi\epsilon_0}\int_0^\Omega dx\;(\Omega-x) {\ln(\Omega)\over
4 R_s}\nonumber\\
&\approx& {\omega^2\over 8\pi E_F} \ln|\omega|
\end{eqnarray}
Thus, the single-particle contribution to the imaginary part of the
self-energy is on the order of $\omega^2 |\ln|\omega||$, and in the
small $|\omega|$ limit, is the dominant term for
${\rm Im}[\Sigma(k_F,\omega)]$.

\subsection{${\rm Im}[\Sigma](k_F,\omega)$ for an impure system}

In this subsection, we show that in a dirty system, the
main contribution to ${\rm Im}[\Sigma(k_F,\omega)]$ comes from the small
$Q$ and $y$ region in Eq.\ (\ref{D8}), which gives a linear contribution
in $\omega$.

Defining $x = (Q - Q_-(y))/y$, we rewrite Eq.\ (\ref{D8})
in the limit of small $\Omega$ as
\begin{equation}
{\rm Im}[\tilde\Sigma(k_F,\Omega)] \approx
\int_0^\Omega dy\ y\int_0^\infty dx \;
{{\rm Im} \Bigl[\epsilon^{-1}\Bigl(y(x + {1\over 2}), y-\Omega\Bigr)\Bigr]
\over y(x+ {1\over 2})}
{1+2x\over 4 \sqrt{x+x^2}}.
\label{D20}
\end{equation}
We first calculate the
inner integral of the right hand side of Eq.\ (\ref{D20}),
\begin{equation}
I(y) = \int_0^\infty dx \;
{{\rm Im} \Bigl[\epsilon^{-1}\Bigl(y(x + {1\over 2}), y-\Omega\Bigr)\Bigr]
\over y(x+ {1\over 2})}
{1+2x\over 4 \sqrt{x+x^2}}.
\end{equation}
We approximate the dielectric function $\epsilon(Q,\Omega)$
by using the diffusive form of the polarizability,
written as an expansion in powers of
$Q$ and $\Omega$ of the numerator and denominator of the Mermin
formula\cite{mermin}. The diffusive form
is strictly valid only for $Q,\Omega << \tilde\gamma$.
However, since the contribution to ${\rm Im}[\Sigma(k_F,\Omega)]$
from the small $Q$ region goes as $\Omega$, and the
contribution from the large $Q$ region goes as $\Omega^2$ both
in the diffusive and the ballistic ({\em i.e.}, $\gamma = 0$) forms,
the error introduced
by using the diffusive form at large $Q$ is insignificant.
The diffusive form of the polarizability is
\begin{eqnarray}
\tilde\Pi_\gamma(Q,\Omega') &\approx& -{ Q^2\over Q^2 -
i\tilde\gamma\Omega/2}\nonumber\\
{\rm Re}\tilde\Pi_\gamma(Q,\Omega')
&\approx& -{Q^4\over Q^4 + (\tilde\gamma\Omega/2)^2}\\
{\rm Im}\tilde\Pi_\gamma(Q,\Omega') &\approx&
-{Q^2\tilde\gamma\Omega\over 2(Q^4 +
(\tilde\gamma\Omega/2)^2}\nonumber
\end{eqnarray}
and hence
\begin{eqnarray}
{{\rm Im}[\epsilon^{-1}(Q,\Omega')]\over Q}
&=&{R_s {\rm Im}[\tilde\Pi]\over (Q-R_s{\rm Re}[\tilde\Pi])^2 +
(R_s{\rm Im}[\tilde\Pi])^2}\nonumber\\
&\approx& {2 R_s\over\tilde\gamma\Omega}
\biggl[1+ \Bigl({2 R_s Q\over\tilde\gamma \Omega}\Bigr)^2\biggr]^{-1}
\end{eqnarray}
where we have ignored terms higher order in $Q,\Omega$.
The inner integral is therefore
\begin{eqnarray}
I(y) &=& {r_s\over 2 \tilde\gamma\Omega} \int_0^\infty dx\ {2x + 1\over
(1 + A_y (2x+1)^2) \sqrt{x^2 + x} }\nonumber\\ &&\\
A_y &=& \Bigl({r_s y\over\tilde\gamma (y-\Omega)}\Bigr)^2.\nonumber
\end{eqnarray}
By a change of variables $z^2 = x^2 + x$, we obtain
\begin{eqnarray}
I(y)
&=& {r_s\over\tilde\gamma\Omega} \int_0^{\infty}
{dz\over 1 + A_y(1+4 z^2)}\nonumber\\
&=& {r_s\over 4\pi \sqrt{A_y(A_y+1)} \tilde\gamma\Omega},
\end{eqnarray}
and hence
\begin{eqnarray}
\tilde\Sigma &\approx& \int_0^\Omega dy\;y I(y)\nonumber\\
&=& {1\over 4\pi\Omega}\int_0^\Omega dy {(y-\Omega)^2\over
\sqrt{(r_s y^2/\tilde\gamma) + (y-\Omega)^2}}\nonumber\\
&=& \Omega \int_0^1 {d\zeta \over 4\pi} { (1-\zeta)^2 \over
\sqrt{(r_s\zeta^2/\tilde\gamma) + (1-\zeta)^2}}.
\end{eqnarray}
The integral is just a function  of $r_s/\tilde\gamma$, and therefore
this shows that $|{\rm Im}[\Sigma(k_F,\omega)]| \sim |\omega|$ in
a dirty system.

\appendix{Estimation of the coupling parameter ${\lowercase{g}}$\\
for a quantum wire}
\label{app:appe}

In this appendix, we estimate the strength of the Tomonaga-Luttinger
coupling parameter $g$
within a quantum wire of width $a$ which is screened by another quantum wire
which is adjacent and parallel to it, a distance $l$ away.
We denote the wires by numbers $1$ and $2$.
We assume that the electrons do not tunnel and are screened
within the random phase approximation.  The effective intrawire
interaction ({\em i.e.}, electrons within the same wire) is given by the
the bare interaction screened by the motion of the electrons in the
adjacent wire.  The Dyson's equation for the screened interaction is
\begin{eqnarray}
V_{\rm eff}(q,\omega)
&=& V_{c,11}(q) + V_{c,12}(q)\Pi_{2}(q,\omega)V_{c,21}(q)\nonumber\\
&&\qquad\qquad
+V_{c,12}(q)\Pi_{2}(q,\omega)V_{c,22}(q)\Pi_{2}(q,\omega)V_{c,21}
+ ... \nonumber\\
&=& V_{c,11}(q) + V_{c,12}\Pi_{2}(q,\omega)(q)V_{c,12}(q)
\sum_{n=0}^\infty \Pi_{2}(q,\omega) V_{c,22}(q)\\
&=& V_{c,11}(q) + {V_{c,12}(q)\Pi_{2}(q,\omega)V_{c,21}(q)\over
1-V_{c,22}(q)\Pi_{2}(q,\omega)},\nonumber
\end{eqnarray}
where $V_{c,ij}$ is the bare Coulomb interaction between electrons
in wires $i$ and $j$, and $\Pi_i$ is the polarizability of wire $i$.

Now assume both wires are identical, so that the intrawire interaction
$V_{11} = V_{22} = V_{a}$, the interwire interaction
$V_{12} = V_{21} = V_{e}$ and
$\Pi_{11}= \Pi_{22}= \Pi$.  Since we are interested in the physics
close to the Fermi surface, we take the $q\rightarrow 0$ asymptotic
forms of these potentials.
The intrawire potential $V_a$ is given by Eq.\ (\ref{defvcq}),
and the interwire
potential $V_e$ is given, in the limit of small $q$, by
\begin{eqnarray}
V_{e}(q) &=& {2e^2\over\epsilon_0}\;
\int_{-a/2}^{a/2} dy
\int_{l-a/2}^{l+a/2} dy' K_0(|q(y-y')|) \cos^2\Bigl({\pi y\over a}\Bigr)
\cos^2\Bigl({\pi y'\over a}\Bigr)\nonumber\\
&\approx& K_0(|ql|)\approx V_{a}(q) - {2e^2\epsilon_0^{-1}} \ln(l/a).
\end{eqnarray}
This gives a long wavelength effective intrawire interaction  of
\begin{equation}
V_{\rm eff}(q,\omega) = {V_{a}(q) - 2e^2\epsilon_0^{-1}\ln(l/a)
\Bigl(2V_a(q) - 2e^2\epsilon_0^{-1}\ln(l/a)\Bigr)\Pi(q,\omega)
\over 1-\Pi(q,\omega)V_a(q)}.
\label{F3}
\end{equation}
In the limit $q\rightarrow 0$, $V_a(q)\rightarrow \infty$ and
$\Pi(q,\omega=0) = -(2/\pi v_F)$, which yields, from Eq.\ (\ref{F3}),
\begin{equation}
V_{\rm eff}(q\rightarrow 0,0) = {v_F\pi\over 2} \Bigl(1 + {8e^2
\over\epsilon_0 v_F\pi} \ln(l/a)\Bigr).
\end{equation}
For $l/a = 10$, and for the parameters of Fig.\ \ref{fig3},
${8e^2 /(\epsilon_0 v_F\pi)}=2.8$, we obtain
$g = 2 V_{\rm eff}/(\pi v_F) = \Bigl(1 + 2.8 \ln(10)\Bigr) \approx 7.5$.
and therefore, using Eq.\ (\ref{31a}), $\alpha\approx 0.15$.

\appendix{${\rm I{\lowercase{m}}}[\Sigma({\lowercase{k}}_F,\omega)]$
as $\omega\rightarrow 0$,\\ in the $GW\Gamma$ and $RPA$ approximation}
\label{app:appf}

In the $GW\Gamma$ approximation,\cite{ting,mahanser} the
Matsubara expression for the self-energy is
\begin{equation}
\Sigma(k,i\nu_n) = -{1\over\beta} \sum_{i\nu_m}
\int {d q\over 2\pi}\; {v_q\over \epsilon(q,i\nu_m)}
\Gamma(q,i\nu_n) {1\over i\nu_n + i\nu_m - \xi_{k+q}}
\label{E1}
\end{equation}
where
\begin{eqnarray}
\epsilon(q,i\nu_n) &=&
1 - V_c(q)\Pi(q,\omega)\Gamma(q,i\nu_n),\nonumber\\
& & \label{E2}\\
\Gamma(q,i\nu_n) &=& 1/[1 + G(q)V_c(q)\Pi(q,\omega)].\nonumber
\end{eqnarray}
$\Gamma$ gives contributions from the vertex corrections.
In the $GW$ and $RPA$ approximations we have used, $\Gamma=1$.
We now show that when vertex
corrections in the form given in Eq.\ (\ref{E2}) are included,
the main conclusions of our calculations regarding the existence
of Fermi surface do not change.

Substituting Eq.\ (\ref{E2}) in Eq.\ (\ref{E1}), one can rewrite
the expression for the self-energy as
\begin{equation}
\Sigma(k,i\nu_n) = -{1\over\beta}\sum_{i\nu_m}\int {d q\over 2\pi}\;
{V_c(q)\over 1-V_c(q)\Pi(q,\omega)[1-G(q)]} {1\over i\nu_n + i\nu_m -
\xi_{k+q}}.
\end{equation}
Physically, this equation says that the main effect that the
vertex corrections have is that the local-field factor $G(q)$
(which accounts for the short-range correlation of electrons beyond
RPA) is to renormalize the dielectric function, changing the
dispersion at large $q$.

One can make the equivalent of the Hubbard approximation\cite{hubbard,singwi}
for $G(q)$ in one dimension to obtain\cite{gold1}
\begin{equation}
G(q) = {1\over 2}{V_c(\sqrt{q^2 + k_F^2})\over V_c(q)}.
\label{E4}
\end{equation}
In the limit where $q\rightarrow 0$, $G(q)\sim |\ln|qa||^{-1}
\rightarrow 0$, and therefore, the local-field correction term
in Eq.\ (\ref{E4}) is negligible for small $q$. However, the behavior
of ${\rm Im}[\Sigma(k_F,\omega)]$ for small $\omega$ only depends on the
integrand of Eq.\ (\ref{E1}) at small $q$, which the
vertex corrections leave essentially unchanged.
Thus the inclusion of vertex corrections of the form Eq.\ (\ref{E2})
does not change $\Sigma(k_F,\omega)$ at small $\omega$,
and hence does not change the conclusions regarding the Fermi surface.

\appendix{Expression for the finite-temperature\\
second-order self-energy}
\label{app:appg}

Here, we give the expression obtained for the second order (in the
screened interaction $w$) self-energy (Fig.\ \ref{fig1}(c))
using our contour deformation method.  We obtained
\begin{equation}
\Sigma(k,z) = \int {d{\bf q}\over (2\pi)^d} \int {d{\bf q'}\over
(2\pi)^d} H_{{\bf k},{\bf q},{\bf q'}}(z)
\end{equation}
where
\begin{eqnarray}
H_{{\bf k},{\bf q},{\bf q'}}(z)
&=& T^2\sum_{i\omega_n}\sum_{i\omega_{n'}}
{w(q',i\omega_{n'}) w(q,i\omega_n)\over
(i\omega_n + z - \xi_{{\bf k}+{\bf q}})(i\omega_n + i\omega_{n'} +
z - \xi_{{\bf k}+{\bf q}+{\bf q'}})
(i\omega_{n'} + z -\xi_{{\bf k}+{\bf q'}})}\nonumber\\
&+& T\sum_{i\omega_n} {w(q,i\omega_n)
w(q',\xi_{{\bf k}+{\bf q'}}-z) R_z(\xi_{{\bf k}+{\bf q'}})
\over (i\omega_n + \xi_{{\bf k}+{\bf q'}} -\xi_{{\bf k}+{\bf q}+{\bf q'}})
(i\omega_n + z -\xi_{{\bf k}+{\bf q}})}\nonumber\\
&+& T\sum_{i\omega_n} {w(q',i\omega_n) w(q,\xi_{{\bf k}+{\bf q}}-z)
R_z(\xi_{{\bf k}+{\bf q}})
\over (i\omega_n + \xi_{{\bf k}+{\bf q}} -\xi_{{\bf k}+{\bf q}+{\bf
q'}}) (i\omega_n + z - \xi_{{\bf k}+{\bf q'}})}\nonumber\\
&-& T\sum_{i\omega_n} {w(q',i\omega_n)w(q,
\xi_{{\bf k}+{\bf q}+{\bf q'}}-z-i\omega_n) R_z(\xi_{{\bf k}+{\bf q}+{\bf q'}})
\over (i\omega_n + z -\xi_{{\bf k}+{\bf q'}})
(i\omega + \xi_{{\bf k}+{\bf q}} - \xi_{{\bf
k}+{\bf q}+{\bf q'}})}\nonumber\\
&+& {w(q,\xi_{{\bf k}+{\bf q}}-z) w(q',\xi_{{\bf k}+{\bf q'}}-z)
R_z(\xi_{{\bf k}+{\bf q}}) R_z(\xi_{{\bf k}+{\bf q'}})
\over \xi_{{\bf k}+{\bf q}} + \xi_{{\bf k}+{\bf q'}}
- \xi_{{\bf k}+{\bf q}+{\bf q'}} -z}\nonumber\\
&-& {w(q',\xi_{{\bf k}+{\bf q'}}-z) W(z,\xi_{{\bf k}+{\bf q}+{\bf q'}}
-\xi_{{\bf k}+{\bf q'}}) R_z(\xi_{{\bf k}+{\bf q'}})n_F(\xi_{{\bf
k}+{\bf q}+{\bf q'}})\over
\xi_{{\bf k}+{\bf q}} + \xi_{{\bf k}+{\bf q'}}
- \xi_{{\bf k}+{\bf q}+{\bf q'}} -z}\nonumber\\
-\int_{-\infty}^\infty &{d\omega\over 2\pi}& {B(q,\omega) w(q',\xi_{{\bf
k}+{\bf q}+{\bf q'}}-z-\omega)
R_z(\xi_{{\bf k}+{\bf q}+{\bf q'}}-\omega) [n_F(\xi_{{\bf k}+{\bf
q}+{\bf q'}}) + n_B(\omega)] \over
(\omega + \xi_{{\bf k}+{\bf q'}}
- \xi_{{\bf k}+{\bf q}+{\bf q'}})(\omega+z -\xi_{{\bf k}+{\bf q}})},
\end{eqnarray}
where
$R_z(E) = n_B(E-z) + n_F(E)$.

\begin{table}
\caption{The small $|\omega|$ forms of the real and imaginary parts of
$\Sigma$ in one and two dimensions, in the $GW$ and $RPA$
approximations, for clean and dirty systems}

\begin{tabular}{||l|c|c||}\hline
         & $d=1$ & $d=2$ \\ \hline
$|{\rm Im}[\Sigma]|$, clean, (plasmon contribution) & \phantom{dunno}
$\omega \sqrt{|\ln(|\omega|)|}$\phantom{dunno} &
\phantom{won't}$\omega^2 $ \phantom{center}\\ \hline
$|{\rm Im}[\Sigma]|$, clean, (single-particle contrib.) &
$\omega/[\ln(|\omega|)]^2$ & $\omega^2|\ln(|\omega|)|$ \\ \hline
Re$[\Sigma]$, clean & $\omega\ln(\omega)$ & $\omega$ \\ \hline
$|{\rm Im}[\Sigma]|$,
dirty & $\omega^2|[\ln(|\omega|)]|^3$ & $|\omega|$ \\ \hline
Re$[\Sigma]$, dirty & $\omega$ & $>\omega$ \\ \hline
\end{tabular}
\label{table1}
\end{table}
\bigskip\bigskip

\begin{table}
\caption{
Many-body characteristics of one-dimensional quantum wire with and without
scattering, in the random phase approximation.}
\begin{tabular}{||l|c|c||}\hline
                & $\gamma = 0$ & $\gamma \ne 0$ \\ \hline
{Fermi liquid?} &  No\ \ $(Z_{F} = 0)$  & Yes\ \ $(Z_{F} \ne 0)$\\ \hline
Re[$\Sigma(k_F,\omega)]$ as $\omega\rightarrow 0$ &
$\sim\;  \omega \log(|\omega|)$  & $\sim\; -\omega$\\ \hline
$|{\rm Im}[\Sigma(k_F,\omega)]|$ as $\omega\rightarrow 0$ &
$\sim\; |\omega|\, |\log(|\omega|)|^{1/2}$
& $\sim\;\omega^2 |\log(|\omega|)|^3$ \\ \hline
$A(k_F,\omega)$ as $\omega\rightarrow 0$ &
$\sim\; \Bigl(|\omega|\, |\log(|\omega|)|\Bigr)^{-1}$ &
$\sim\;|\log(|\omega|)|^3$|\\ \hline
{Inelastic scattering rate $\Gamma$} &
$(k-k_c)^{-1/2}$  as $k \rightarrow k_c^+$ &
Finite for all $k$ \\ \cline{2-3}
& $0$ below plasmon emission threshold
&  Nonzero except at $k_F$  \\ \hline
\end{tabular}
\label{table2}
\end{table}

\figure{(a) Electron self-energy in leading order
in the effective dynamical screened Coulomb interaction. The straight line
denotes the bare Green's function and the thick wavy line denotes
the dynamically screened interaction. (b) Dyson's equation
for the screened interaction, within the random phase approximation
(the thin wavy line is the bare Coulomb interaction).
Impurity effects are introduced in the electron-hole polarization
bubble through the Mermin formula.  (c) Second order diagram generally ignored
in this paper.  In appendix \ref{app:appb},
we show that this diagram, for $k=k_F$ and
$\omega=-\xi_{k_F}$ with a screened interaction which is a delta-function
in real-space ({\sl i.e.} a constant in momentum space), is identically zero.
\label{fig1}}

\figure{Regions of $q-\omega'$ space
with non-zero ${\rm Im}[\epsilon^{-1}(q,\omega')]$.
The grey shaded regions correspond to the single-particle excitations,
${\rm Im}[\epsilon] \ne 0$, and the thin line
corresponds to the plasmon dispersion ${\rm Re}[\epsilon] = 0$.
To evaluate ${\rm Im}[\Sigma(k,\omega)]$ within the GW approximation,
one integrates the function $V(q){\rm Im}[\epsilon^{-1}(q,\omega')]$
over the two bold line segments, which belong to the curve
$\omega'(q) = \omega - \xi_{k+q}$) [see Eq. (\ref{imsigma})].
The figure shows that both the single-particle excitation region and
the plasmon line contribute to ${\rm Im}[\Sigma(k_F,\omega)]$.
\label{fig2}}

\figure{ (a),(b) Real (bold lines) and imaginary
(thin lines) parts of the self-energy $\Sigma(k,\omega)$ and
(c), (d) spectral functions $A(k,\omega)$ as functions of the
energy $\omega$, for $\gamma= 0$ (solid lines)
and $\gamma = E_F$ (dashed lines).
Figs.\ (a) and (c) are for $k=0$, and (b) and (d) are for
$k=k_F$.
The parameters
used are $k_F a = 0.9$ and $r_s = (4 m_e e^2/\pi \hbar^2 k_F
\epsilon_{0}) =  1.4$,\cite{define}
(corresponding to $a = 100\,{\rm \AA}$, a density of
$0.56 \times 10^6 \, {\rm cm}^{-1}$ and a Fermi energy of
$E_F \approx 4.4\;{\rm MeV}$ in GaAs).
The vertical arrows in (c) denote $\delta$-functions in the
spectral function at $\omega = -4.9, 0.9$
with weights $(2\pi)\times 0.32 $ and $(2\pi)\times 0.33$, respectively.
Note that in (d), $A(k_F,\omega)$
has a $\delta$-function of weight $(2\pi) \times 0.3$
at $\omega=0$ for $\gamma=E_F$, but not
for $\gamma = 0$.  The straight lines are given by $\omega-\xi_k-\mu$,
and their intersections with ${\rm Re}[\Sigma]$ indicate the solutions
to Dyson's equation and correspond to a quasiparticle peak.\label{fig3}}

\figure{
(a) Momentum distribution function $n_k$, around $k/k_F = 1$,
(b) Fermi surface
renormalization factor $Z_F$, and (c) density of states $\rho(\omega)$
for small $\omega$, of a quasi-one dimensional electron gas
for $k_F a = 0.9$ and $r_s =  1.4$ (as in Fig.\ \ref{fig3}),
for various impurity scattering rates $\gamma$.
The bold lines refer to $k > k_F$, and the thin lines to $k < k_F$.
For $\gamma = 0$, $n_k$ is continuous at $k = k_F$, implying that the
system is non-Fermi liquid, but for $\gamma\ne 0$
a discontinuity occurs at $k = k_F$, signalling the
presence of a Fermi surface. The magnitude of the discontinuity in $n_{k_F}$
is given by $Z_F$.\label{fig4}}

\figure{(a) Inelastic scattering rates $\Gamma(k)$,
and (b) the corresponding mean free path $l(k) = \Gamma(k)k/m$,
as a function of $k$, for various  $\gamma$'s (electron-impurity
scattering rates), for $k_F a = 0.9$ and $r_s = 1.4$.
Within RPA, for $\gamma = 0$, the $\Gamma(k)$ is identically
zero below $k = k_c$ because energy and momentum conservation
prohibits single particle excitations and plasmon emissions.
Above $k_c$, the scattering rate is caused by plasmon emissions.
For $\gamma \ne 0$, the plasmon line broadens and momentum conservation
is relaxed, resulting in a nonzero $\Gamma $ for $k < k_c$.
\label{fig5}}

\figure{Total bandgap renormalization (${\rm Re}
[\Sigma_e + \Sigma_h]$ at $k = 0$, $\omega = \xi_{k=0}$)
as a function of electron density in the quantum wire
for various wire widths with parameters corresponding to GaAs.
\label{fig6}}

\figure{Electron self-energy for impurity scattering,
to lowest nontrivial order in the impurity potential,
used to calculate the momentum
distribution function of a disordered one-dimensional electron gas (see
Fig.\ \ref{fig8}).  The dashed lines are the screened impurity scattering
potential, which we approximate by $\delta$-functions in real space.
\label{fig7}}

\figure{(a) The momentum distribution function,
for $\gamma_{\rm imp}/E_F = 0.02,0.2$ and $2$,
and (b) spectral function (bold solid line), and
real (thin solid line) and imaginary (dashed line) parts of the
self-energy and for $k=k_F$, $\gamma_{\rm imp} = 2 E_F$,
of a disordered non-interacting
one-dimensional electron gas at $T=0$, calculated with the self-energy
shown in Fig.\ \ref{fig7}.
$\gamma_{\rm imp} = 2 m N_i U_0^2/k_F$ is
the Born approximation impurity scattering rate for the electron
of energy $E_F$ (the Fermi energy corresponding to the pure system of
the same electron density).
The inset in Fig.\ (a) shows the chemical potential $\mu$ for fixed density,
as a function of the disorder.   In Fig.\ (b),
the arrows correspond to delta functions.
The dashed arrow marks the position of the delta function for
the pure system, and has weight $2\pi$. The solid arrow has
weight $(2\pi)\times 0.39$.
The straight dotted line is given by $\omega - \xi_k - \mu$, and its
intersections with ${\rm Re}[\Sigma]$ correspond to a quasiparticle
peaks.\label{fig8}}

\figure{The distribution function of a Luttinger
liquid with Fermi surface exponent $\alpha=0.15$, which corresponds to
a dimensionless coupling constant $g\approx 7.5$, for various scattering rates.
The Luttinger model linearizes the dispersion around the Fermi surface,
and therefore the curves strictly should only be valid around $k = k_F$.
\label{fig9}}

\figure{The contour of integration
${\cal C}$ for Eq. (\ref{contour}).
The hatched real axis indicates a branch cut due to
$w({\bf q},\omega)$ in the integrand of Eq.\ (\ref{contour}).
The crosses mark the poles
due to the integrand;
the ones on the imaginary axis are due to $n_B(\omega)$,
and the isolated pole is due to the denominator.  The residues of the
poles on the imaginary axis give $h_{{\bf k},{\bf q}}(z)$,
while the residue of the
isolated pole gives $\tilde h_{{\bf k},{\bf q}}(z)$.\label{fig10}}

\figure{(a), (b) Real and (c), (d) imaginary
parts of the self-energies and
(e), (f) spectral functions, as a function of frequency,
for electrons in a quantum wire, for $k=0,k_F$, at finite temperatures.
As in previous figures, $k_F a = 0.9$ and $r_s = 1.4$,
which correspond to $E_F \approx 4.4\,{\rm meV}
\approx 50\,{\rm K}$ in GaAs.
The discontinuities is ${\rm Im}[\Sigma(k,\omega)]$ at $T=0$,
which arise from virtual plasmon emission thresholds,
broaden with increasing temperature because the plasmon peaks broaden
due to Landau damping.  The logarithmic divergence in the $T\ne 0$
${\rm Im}\Sigma(k,\omega)$ at $\omega = E_k - \mu_0(T)$,
due to the divergence of the one-dimensional Born-approximation
scattering rate, shifts slightly with respect to temperature because
of the temperature dependence of $\mu_0$, the non-interacting
chemical potential.\label{fig11}}

\figure{Temperature dependence of the band-gap
renormalization due to conduction electrons for a wire width of
$100\;{\rm \AA}$ in GaAs, for electron densities of $10^4\; {\rm cm}^{-1}$
(solid lines), $10^5\;{\rm cm}^{-1}$ (dotted lines) and $10^6\;{\rm
cm}^{-1}$ (dashed lines).
The thin lines are for the electrons
(${\rm Re}[\Sigma_{\scriptscriptstyle{\rm electron}}(k=0, \xi_{k=0})$])
the light bold lines are for the
holes (${\rm Re}[\Sigma_{\scriptscriptstyle{\rm hole}}
(k=0,\xi_{k=0})]$ and the heavy bold lines are for the sum of the two.
The densities $n = 10^{4}\;{\rm cm}^{-1}$, $10^5\;{\rm cm}^{-1}$ and
$10^6\;{\rm cm}^{-1}$ correspond to Fermi temperatures of
$E_F = 1.6\times 10^{-2}\;{\rm K}$, $1.6\;{\rm K}$ and
$160\;{\rm K}$, respectively. \label{fig12}}

\figure{(a) A schematic of the band diagram of a
one-dimensional tunneling hot electron transistor,
where electrons are injected from
the emitter into the base (which contains a Fermi sea of electrons)
and (b) energy-- vs.\ momentum--loss diagram for the injected
electron.  In the transistor, the electrons that are scattered
below the base--emitter barrier do not reach the collector, and
therefore the fraction of the injected electrons reaching the collector
depends on the electron scattering rate in the base region.
In Fig.\ (b), the solid (dashed) line indicate the energy- vs.\
momentum-loss for electrons injected
into the base region below (above) the plasmon emission threshold
({\em i.e.},
the solid line is for $k < k_c$ and the dashed line is for $k > k_c$).
The intersections of the energy-- vs.\ momentum--loss curve and the
plasmon dispersion curve (bold line) indicates the wavevectors
at which plasmons are emitted;
if there is no intersection (as with the solid line),
plasmon emissions are not allowed.
As the energies of the injected electrons is raised above the plasmon
emission threshold, the scattering rate increases dramatically (see
Fig.\ \ref{fig14}), drastically reducing the fraction of
injected electrons that
reach the collector.\label{fig13}}

\figure{(a) Momentum scattering rate $\Gamma_{{\rm t},k}$ and (b)
the corresponding mean free path, $l_k = v_k/\Gamma_{{\rm t},k}$.
of an electron in a doped one-dimensional quantum wire,
as a function electron momentum, for various temperatures.
As in previous figures, $k_F a = 0.9$ and $r_s = 1.4$.\label{fig14}}

\figure{Calculated one-dimensional plasmon
dispersion in a GaAs quantum wire within RPA for (a) $T=0$ and
various $\gamma$'s, (b) $\gamma=0$ and various temperatures, calculated
by finding the zeros of $\epsilon(q,\omega)$ on the complex frequency
plane.  The curves with $\omega > 0$ give the real part
and those with $\omega < 0$ give the imaginary part.
The curves include full $V_c(q)$ for infinite square well confinement.
As in previous figures, the system parameters are $ k_F a = 0.9$ and
$r_s = 1.4$,  which, for GaAs, correspond
to a Fermi wavevector of $k_F = 0.88\times 10^{6}\,{\rm cm}^{-1}$.
In (b), we show both the long wavelength result, $\omega_{\rm p}^2 =
n q^2 V_c(q)/m$, using $V_c(q) =
2e^2|\ln(qa)|/\kappa$ (dotted line) and the
full $V_c(q)$ (long dashed lines).
Experimental results of reference \cite{goni}
compared with RPA theory (of reference \cite{li}) are shown
as an inset in Fig.\ (a).\label{fig15}}

\figure{Regions of integration for the calculation of the
second-order self-energy, for $k=k_F$ and $\omega=0$.\label{fig16}}


\begin{references}
\bibitem{define} This definition of $r_s$ (which is more convenient)
differs by a factor of 2 from the definition used earlier in
references \cite{hu}, \cite{huappl} and \cite{hutobep}.
\bibitem{problem} For numerical reasons, it was difficult to obtain
curves for $\gamma/E_F$ much smaller than $10^{-2}$.
\bibitem{rigour} Rigorously, one should prove that
the integral over ${\bf q}$ does not affect the analyticity;
for the $H_{{\bf k},{\bf q}}(z)$ in this case, it does not.
\bibitem{enhancement} We should point out that it is claimed
that for short-ranged interactions (which would be the case if the
intrawire interactions are strongly screened by external metallic gates
or other wires), there is an enhancement in the absorption
cross section.  See reference \cite{ogawa}.
\bibitem{expand} This expansion breaks down when $C(Q)\Omega/\Gamma \ll 1$,
but since $C(Q)$ diverges only logarithmically as $Q\rightarrow 2$, the
size of the region in the integral where the expansion breaks down is
$\sim\exp(-1/\Omega)$, and can be ignored as $\Omega\rightarrow 0$.
\end{references}
\end{document}